\documentclass[twoside,leqno]{article}

\usepackage[letterpaper]{geometry}

\usepackage{siamproceedings}

\usepackage[T1]{fontenc}
\usepackage[utf8]{inputenc} 
\usepackage{utf8math}

\usepackage{inconsolata}
\usepackage{mathrsfs}
\usepackage{mathtools} % includes amsmath
\usepackage{amssymb}
\usepackage{microtype}
\usepackage{amscd}
\usepackage{thm-restate}
\usepackage{enumerate}
\usepackage{framed}
\usepackage{bm}
\usepackage{dirtytalk}
\usepackage{mdframed}
\usepackage{hyperref}
\usepackage{cleveref}

\newif\ifcomments

\commentstrue
\ifcomments
 
\usepackage{todonotes}

 \newcommand{\fnote}[2][]{\todo[#1,backgroundcolor=cyan!20!white,bordercolor=black,linecolor=cyan,size=\small]{Fritz: #2}}
 
 \newcommand{\dnote}[2][]{\todo[#1,backgroundcolor=orange!20!white,bordercolor=blue,linecolor=green,size=\small]{Daniel: #2}}

\else
 
\usepackage[disable]{todonotes}

 \newcommand{\fnote}[2][]{\todo[#1,backgroundcolor=cyan!20!white,bordercolor=black,linecolor=cyan,size=\small]{Fritz: #2}}
 
 \newcommand{\dnote}[2][]{\todo[#1,backgroundcolor=orange!20!white,bordercolor=blue,linecolor=green,size=\small]{Daniel: #2}}

\fi

\newcommand{\F}{\mathbb{F}}
\newcommand{\R}{\mathbb{R}}

\newcommand{\C}{\mathbb{C}}

\newcommand{\cM}{\mathcal{M}}

\newcommand{\supp}{\mathrm{supp}}

\newcommand{\mA}{\mathbf{A}}

\newcommand{\mB}{\mathbf{B}}

\newcommand{\mV}{\mathbf{V}}
\newcommand{\mD}{\mathbf{D}}
\newcommand{\mM}{\mathbf{M}}

\newcommand{\T}{\mathsf{T}}

\DeclareMathOperator{\si}{si}

\DeclareMathOperator{\dist}{dist}
\DeclareMathOperator{\rank}{rank}

\DeclareMathOperator{\aff}{aff}

\DeclareMathOperator{\spa}{span}
\DeclareMathOperator{\lines}{lines}
\DeclareMathOperator{\maxlines}{maxlines}
\DeclareMathOperator{\Tr}{Tr}

\newcommand{\CCC}{{\mathbb C}}

\newcommand{\QQ}{{\mathbb Q}}
\newcommand{\RR}{{\mathbb R}}
\newcommand{\ZZ}{{\mathbb Z}}

\title{Excluding a Line Minor via Design Matrices and Column Number Bounds for the Circuit Imbalance Measure}
\date{}

\author{Daniel Dadush\thanks{Centrum Wiskunde \& Informatica and Utrecht University, The Netherlands. Email: {\tt dadush@cwi.nl}. Supported by ERC Starting Grant no. 805241-QIP.} \and Friedrich Eisenbrand\thanks{EPFL, Switzerland. Email: {\tt friedrich.eisenbrand@epfl.ch}. Supported by Swiss National Science Foundation (SNSF) grant 10000183.} \and Rom Pinchasi\thanks{Technion, Israel. Email: {\tt room@math.technion.ac.il}.} \and Thomas Rothvoss\thanks{University of Washington, Seattle. Email: {\tt rothvoss@uw.edu}. Supported by NSF grant 2318620 \emph{AF: SMALL: The Geometry of Integer Programming and Lattices}.} \and Neta Singer\thanks{EPFL, Switzerland. Email: {\tt neta.singer@epfl.ch.}}}

\begin{document}
\maketitle

%\fancyfoot[R]{\scriptsize{Copyright \textcopyright\ 2026 by SIAM\\
%Unauthorized reproduction of this article is prohibited}}

\begin{abstract}
For a real matrix $\mA \in \R^{d \times n}$ with non-collinear columns, we show
that $n \leq O(d^4 \kappa_\mA)$ where $\kappa_\mA$ is the \emph{circuit imbalance
measure} of $\mA$. The circuit imbalance measure $\kappa$ is a real analogue of
$\Delta$-modularity for integer matrices, satisfying $\kappa_\mA \leq \Delta_\mA$
for integer $\mA$. The circuit imbalance measure has numerous applications in the context of linear
programming (see Ekbatani, Natura and V{\'e}gh (2022) for a survey). Our
result generalizes the $O(d^4 \Delta_\mA)$ bound of Averkov and Schymura (2023)
for integer matrices and provides the first polynomial bound holding for all
parameter ranges on real matrices.

To derive our result, similar to the strategy of Geelen, Nelson and Walsh (2021) for $\Delta$-modular matrices, we show that real representable matroids induced by 
$\kappa$-bounded matrices are minor closed and exclude a rank $2$ uniform
matroid on $O(\kappa)$ elements as a minor (also known as a line of length
$O(\kappa)$).  

As our main technical contribution, we show that any
simple rank $d$ complex representable matroid which excludes a line
of length $l$ has at most $O(d^4 l)$ elements. This complements the tight
bound of $(l-3)\binom{d}{2} + d$ for $l \geq 4$, of Geelen, Nelson and Walsh
which holds when the rank $d$ is sufficiently large compared to $l$
(at least doubly exponential in $l$).     

Our proof of the above relies on an improvement of a Sylvester-Gallai type
theorem of Dvir, Saraf and Wigderson (2014). Refining their design matrix
technique, we show that for any full dimensional set of $n$ points in $\CCC^d$
there always exists a point that lies on at least $(1-\frac{4}{d})n$ many
distinct lines (the constant $4$ is improved from $12$). The excluded minor
bound follows by inductively applying this result to find good elements to
contract in the matroid, where the improved constant reduces the dependence
on $d$ from $d^{12}$ to $d^4$. Interestingly, by relying on geometric
techniques, our proof avoids the use of any difficult matroid machinery.
\end{abstract}

\section{Introduction}

A fundamental concept seen in combinatorial optimization is the concept of a \emph{totally unimodular (TU)} matrix, which is a matrix  $\mA \in \mathbb{Z}^{d \times n}$ where all square submatrices have determinant $-1$, $0$ or $+1$. The importance of this notion comes from the fact that all polyhedra of the form $P = \{ x \in \mathbb{R}^d \mid \mA^\T x \leq b \}$ with TU matrix $A$ and integral right hand side $b$ have only integral vertices and hence any integer program
\begin{equation}
  \label{eq:1}
  \max\{ c^\T x \mid \mA^\T x \leq b, x \in \mathbb{Z}^d\} 
\end{equation}
can be solved in polynomial time, see e.g. the textbook by
Schrijver~\cite{schrijver1999theory}. The most natural generalization of this
concept is the one of a \emph{totally $\Delta$-modular} matrix $\mA \in
\mathbb{Z}^{d \times n}$ where all square submatrices $\mB$ have $|\det(\mB)|
\leq \Delta$. As a slight alternative, \emph{$\Delta$-modular} matrices have
been considered where this condition only needs to hold for $r \times r$
submatrices where $r$ is the rank of $\mA$. Artmann, Weismantel and
Zenklusen~\cite{artmann2017bimodular} proved that 2-modular integer programs
can be solved in polynomial time. Whether  an integer program~\eqref{eq:1}
can be solved in polynomial time for any fixed constant $\Delta$ is a highly
visible open problem. Fiorini et al.~\cite{fiorini2022integer} have shown
this to be the case if $A$ additionally contains at most two nonzero entries
per column or two nonzero entries per row. 

One property that is indeed known to generalize from the totally unimodular
case is that the number of distinct columns (or non-parallel columns
depending on the context) of such a matrix $\mA \in \mathbb{Z}^{d \times n}$ is
bounded. It has been known since the work of Heller~\cite{heller1957linear}
that the number of columns of a TU matrix is at most $n \leq
\frac{1}{2}(d^2+d)$, which is tight for the network matrix of a complete
graph on $d$ vertices together with an appended identity matrix. For the case
of general $\Delta \geq 2$, the question was first studied by Lee
\cite{lee1989subspaces} who gave a bound of the form $O(d^{2\Delta})$ for a fixed $d$ and $\Delta \rightarrow \infty$.
Glanzer, Weismantel and Zenklusen~\cite{glanzer2018number} gave the first
slightly super-polynomial bound of $\Delta^{\log_2 \log_2 \Delta+2} d^2$.
Polynomial bounds in $\Delta$ and $m$ on the number of \emph{distinct} columns have been provided by  Averkov and
Schymura~\cite{averkov2023maximal}  of $O(d^4\Delta)$
% based on a reduction to the unimodular case.
and by Lee, Paat, Stallknecht and Xu~\cite{lee2023polynomial}
of $(d^2+d)\Delta^2/2$ the latter of which generalizes the bound of Heller for $\Delta = 2$. The work of Geelen, Nelson and
Walsh~\cite{geelen2024excluding} implies a bound of $d^2 + g(\Delta) d$ for
some at least doubly-exponential function $g(\Delta)$. Paat, Stallknecht,
Walsh and Xu~\cite{paat2024column} give a polynomial bound on $g(\Delta)$ and
showed that the number of \emph{non collinear columns} (which enables a slightly
smaller bound than for distinct columns) of a $\Delta$-modular matrix is
bounded by $\binom{d}{2}+80\Delta^7 d$. These bounds are based on deep
matroid techniques. Non collinear columns corresponds to the elements of simple $\Delta$-modular matroids while distinct columns correspond to the elements of an arbitrary $\Delta$-modular matroid. This means that bounds on distinct columns are a factor $\Delta$ stronger than bounds on non collinear columns in general. 

\subsubsection*{Condition Measures for Real Matrices} 
 
In the same way that the study of $\Delta$-modularity for integer matrices
was motivated by integer programming, there has been significant research on
condition measures for real constraint matrices motivated by linear
programming (LP). The goal has in general been to understand how these affect
both the geometry and solvability of the underlying LP. In particular,
condition measures for real constraint matrices have been used to bound the
combinatorial diameter~\cite{bonifas2014sub,dadush2016shadow} and circuit
diameter of polyhedra~\cite{dadush2022circuit}, the iteration complexity of
interior point
methods~\cite{vavasis1996primal,monteiro2005new,dadush2024scaling} and first
order methods~\cite{cole2024first} for solving LPs exactly, as well as
sensitivity and proximity
bounds and their algorithmic applications~\cite{MR861043,guler1995approximations,dadush2020revisiting}. We refer the
reader to the excellent survey of Ekbatani, Natura and
V{\'e}gh~\cite{ekbatani2022circuit} for a thorough overview of these
applications in the context of LP.

We now introduce some basic notation together with well studied condition
measures. For $\mA \in \R^{d \times n}$, $\rank(\mA)$ is the rank of $\mA$, $\|\mA\|_{\rm op}$
denotes the operator norm, $\mA_B$, $B \subseteq [n]$, denotes the columns
indexed by $B$ and $\spa(\mA_B)$ is their linear span. For $x \in \R^n$,
$\supp(x) := \{i \in [n]: x_i \neq 0\}$. We say that $x$ is a \emph{minimal linear dependence}
of $\mA$ if $\mA x=0$ and if $\mA y = 0$, $\supp(y) \subsetneq \supp(x)$ implies $y =
0$. For an additional set $C \subseteq \R^n$, $\dist(x,C) := \inf_{y \in C}
\|x-y\|$ is the Euclidean distance between $x$ and $C$.   

The three condition measures of a real matrix $\mA=(a_1,\dots,a_n) \in \R^{d
\times n}$ for $a_i$ the $i^{th}$ column of $\mA$, which underlie all the applications mentioned above, are as
follows:  

\begin{itemize}
\item Circuit imbalance measure~\cite{ekbatani2022circuit}: \\[1ex]
$\kappa_{\mA} := \max \left\{\left|\frac{x_i}{x_j}\right|: i,j \in [n], x \text{ minimal lin.~dependence of } \mA, x_j \neq 0~\right\}$.
\item Dikin-Stewart-Todd condition measure~\cite{dikin1967iterative,stewart1989scaled,todd1990dantzig}: \\[1ex]
$\bar{\chi}_\mA := \sup \{\|\mA^\T (\mA\mD\mA^\T)^{-1} \mA \mD\|_{\rm op}: \mD \text{ an $n \times n$ positive diagonal }\}$.
\item $\delta$-distance measure~\cite{brunsch2013finding}: \\[1ex]
$\delta_\mA := \min \left\{ \frac{{\rm dist}(a_i,\spa(\mA_{B \setminus \{i\}}))}{\|a_i\|}: i \in B \subseteq [n], \mA_B \text{ a basis of $\mA$ }\right\}$.  
\end{itemize}

While seemingly very different, e.g., measuring ratios of nonzeros in
minimal linear dependencies, the operator norms of oblique projections, and
minimum nonzero distances between a column and spanned hyperplane, it turns
out that these condition measures are tightly related. In particular, we have
the inequalities $\kappa_\mA \leq \bar{\chi}_\mA \leq \sqrt{n}
\kappa_{\mA}$~\cite{vavasis1994stable,dadush2024scaling}, and for a matrix with
unit norm columns $\kappa_\mA \leq 1/\delta_\mA$~\cite{dadush2022finding,ekbatani2022circuit}.
Lastly, if $\mA$ has rank $d$, there exists a basis $\mA_B$ such that
$1/\delta_{\mA_B^{-1} \mA} \leq d \kappa_\mA$~\cite{dadush2022finding,ekbatani2022circuit}.  

Given the many applications listed above, a pressing question is to better
understand the structure of real matrices of bounded condition measure. In
contrast to the in-depth (generally matroidal) study of the structure of
$\Delta$-modular matrices, the research on the structure of such real
matrices has been significantly more limited. 

As with $\Delta$-modularity, the most basic question is how many columns can
a real matrix $\mA \in \R^{d \times n}$ with bounded condition number and fixed
rank $d$ have? For this question to make sense in the real context, one must
now insist on pairwise non-collinear columns instead of distinct columns, since
otherwise the number can trivially be infinite. Of the list of condition
measures above, we will restrict attention in the rest of the paper to
the circuit imbalance measure $\kappa_\mA$. As it is in general a lower bound on
the remaining condition measures, upper bounds on $n$ in terms of $\kappa_\mA$ and $d$ also apply for the other condition measures. Furthermore, it is also easy to show that $\kappa_\mA \leq
\Delta_\mA$ for integer matrices. To see this, we note that by Cramer's rule,
one can define $\kappa_\mA$ in terms of ratios of determinants:
\[
\kappa_\mA := \max \left\{\left|\frac{\det(\mA_{B \cup \{j\} \setminus \{i\}})}{\det(\mA_B)}\right|: \mA_B \text{ a basis of } \mA, j \notin B, i \in B\right\}.
\]
Since the determinant of an integer matrix is always integral, the
denominators above are all at least $1$ in absolute value, and hence the
absolute ratio is always upper bounded by $\Delta_\mA$.

It is also useful to note that $\kappa_\mA$ can be much smaller than
$\Delta_\mA$ even for simple matrices. In particular, if $\mA$ is the
unsigned node arc incidence matrix of a complete graph on $3n$ vertices, then
$\Delta_\mA = 2^n$ (more generally, it is $2$ to the power of the odd-cycle
packing number of the graph), whereas $\kappa_\mA = 2$ in this case.
Interestingly, there is a tight connection when the circuit imbalance measure
is $1$. As shown by Cederbaum~\cite{cederbaum1957matrices} using different
terminology, $\kappa_\mA = 1$ if and only if $\mA$ is row equivalent to a TU
matrix.

\subsection{Our contributions}

As our main contribution, we show that the number of non-collinear columns of a real matrix is bounded by a polynomial in both $\kappa$ and rank. 
 
\begin{theorem}
\label{thm:main-kappa}
If $\mA \in \R^{d \times n}$ for $d \geq 4$ is a matrix with non-collinear columns, then $n \leq \pi d^4 \kappa_{\mA}$.
\end{theorem}

From a quantitative perspective, we match up to a constant factor the $O(d^4
\Delta_\mA)$ bound of Averkov and Schymura for integer matrices (recall that
$\kappa_\mA \leq \Delta_\mA$), though in a significantly more general setting.
Furthermore, the dependence on $\kappa_\mA$ must be linear even for $d=2$. In
particular, it is not hard to show that the rank $2$ matrix
$\mA=(a_0,\dots,a_{n-1})$, $n \geq 2$, with columns $a_i :=
(\cos((i/n)\pi),\sin((i/n)\pi))^\T$, $i \in \{0,\dots,n-1\}$ (i.e.,
evenly spaced on the upper half circle), satisfies $\kappa_\mA = \Theta(n)$. This can be seen by using the determinant ratio definition of $\kappa_\mA$ and taking $\left| \frac{\det\left( a_0, a_{n/2} \right)}{\det(a_0, a_1)}  \right| = \left|\frac{1}{\sin(\frac{\pi}{n})}\right| \geq 
\frac{n}{\pi}$.  \\    

To prove our result, we employ a combination of matroid theoretic and 
geometric techniques. As a starting point, we follow an analoguous strategy
to that of Geelen, Nelson and Walsh~\cite{geelen2024excluding} for
$\Delta$-modular matrices. In the language of matroid theory, they show that
the set of matroids representable by $\Delta$-modular matrices is minor closed
and excludes the rank $2$ uniform matroid on $l := O(\Delta)$ elements, denoted
by $U_{2,l}$ (referred to as a line of length $l$). This allows them to
reduce the question to bounding the number of distinct elements in simple
real or complex representable matroids that exclude a line of suitable length
$l$ as a minor. Our first step is to show that the same statements hold with
$\Delta$-modular replaced with $\kappa$-bounded.         

\begin{lemma}
\label{lem:kappa-minor}
For $\kappa \geq 1$, the set of matroids representable by a real matrix with
circuit imbalance measure at most $\kappa$ form a minor closed family which
excludes a line of length $O(\kappa)$.  
\end{lemma}

For the sake of readers unfamiliar with matroid concepts (see
\Cref{sec:matroid-prelims} for basic definitions),  we restate the excluded
minor condition in simple linear algebraic terms. We recall that the linear
matroid $\cM$ induced by a real matrix $\mA = (a_1,\dots,a_n) \in \R^{d
\times n}$ is the set of columns of $\mA$, and the set of all subsets $S \subseteq [n]$ for which the
columns $\{a_i: i \in S\}$ are linearly independent, that is, $\rank(\mA_S) =
|S|$. The rank of the matroid is the size of the largest independent set which is  $\rank(\mA)$. We say that $\cM$ contains a $U_{2,l}$ minor if there exists disjoint sets $S,T
\subseteq [n]$ with $|T|=l$, $\rank(\mA_S)+2 = \rank(\mA_{T \cup S}) = \rank(\mA_{S \cup \{i,j\}})$, $\forall i,j \in T$, $i \neq j$. Stated
geometrically, letting $\Pi := \Pi_{\spa(\mA_S)^\perp}$ denote the projection
onto the orthogonal complement of $\spa(\mA_S)$, the vectors $\{\Pi(a_i): i
\in T\}$ are nonzero non-collinear vectors which span a $2$ dimensional
subspace. We further extend the above
definitions to linear matroids induced by complex matrices $\mA \in \C^{d
\times n}$, where the span and rank are over $\C$. Matroids that can be
induced by real (complex) matrices are said to be real (complex)
representable.

The minor closedness in \Cref{lem:kappa-minor} is deduced from the fact that minor closed operations on matrices (deletion of columns and contraction) cannot increase
the circuit imbalance measure. This is in essence a matroid theoretic restatement of well
known properties of $\kappa$ in the literature (see for example
\cite[Proposition 3.17]{ekbatani2022circuit}). The exclusion of length $O(\kappa)$
lines as minors then corresponds to showing that \Cref{thm:main-kappa} holds for
the base case $d=2$. This is proved 
%using the constant factor equivalence
%of $\kappa$ and $1/\delta$ in the two dimensional case, followed 
by a simple
packing argument on the circle to lower bound $\kappa$.     

With the above reduction, we now state two main results of Geelen, Nelson and
Walsh bounding the number of elements of simple linear matroids excluding a line, where
simplicity means that any matrix representation has nonzero and
non-collinear columns. Using \Cref{lem:kappa-minor}, by replacing the line
size by $O(\kappa)$ one immediately gets a corresponding bound on the number
of non-collinear columns of a $\kappa$ bounded matrix.  

\begin{theorem}[\cite{geelen2024excluding}] 
\label{thm:real-rep}
For $l \geq 5$, a simple rank $d$ real representable matroid excluding
$U_{2,l}$ as a minor has at most $d^2 + g(l) d$ elements for some function
$g$ of $l$.
\end{theorem}
\begin{theorem}[\cite{geelen2021excluding}] 
\label{thm:complex-rep}
For $l \geq 4$, a simple rank $d \geq g(l)$ complex representable matroid
excluding $U_{2,l}$ as a minor has at most $(l-3)\binom{d}{2} + d$ many
elements, for some function $g$ of $l$.  Moreover, equality holds if and
only if the matroid is isomorphic to the rank $d$ cyclic Dowling geometry of
order $l-3$.
\end{theorem} 

We note that any matroid excluding a $U_{2, 4}$ minor is isomorphic to the binary matroid. The rank $d$ cyclic Dowling geometry of order $t$, yielding the tight example
above, is the linear matroid induced by matrix containing the standard basis
$e_1,\dots,e_d$ together with $e_i - \zeta^k e_j$, $1 \leq i < j \leq d$ and $0
\leq k \leq t-1$, where $\zeta \in \CCC$ is any $t$th root of unity
\footnote{The minors induced by contracting any number of elements in a cyclic
Dowling geometry will be isomorphic, up to removing parallel elements and loops, to a
cylic Dowling geometry of lower rank. Furthermore, a rank $2$ cyclic Dowling
geometry of order $t$ contains exactly $t+2$ non-parallel elements.}.

While the above bounds are very strong, in both cases the function $g(l)$
given by the proofs is at least double exponential in $l$. In particular,
they do not imply bounds that depend polynomially on both the rank and
$\kappa$ in all ranges of parameters.  

Our main technical contribution in this regard, is to give such a polynomial
bound in the complex (and thus also real) setting, complementing the above
results. Combined with \Cref{lem:kappa-minor}, this immediately implies
\Cref{thm:main-kappa}.  

\begin{theorem}
\label{thm:complex-rep-ours}
For $l \geq 2$, a simple rank $d$ complex representable matroid excluding
$U_{2,l}$ as a minor has at most $\frac{1}{4} d^4 l$ elements for $d \geq 3$.  
\end{theorem}

Given the generality of the above result, one may wonder whether the column
bound in \Cref{thm:main-kappa} also extends in the complex setting. Indeed,
there is no difficulty in defining the circuit imbalance measure for complex
matrices, one just replaces absolute value by modulus. Via the exact same
logic as above it can be shown that an $O(d^4 \kappa^2)$ bound holds in this
setting, and that the $\kappa^2$ dependence is unavoidable. In particular,
there are rank $2$ matrices with $\Theta(\kappa^2)$ entries and circuit
imbalance measure $\Theta(\kappa)$. %As the main applications we are aware of are in the real setting, we defer details of the complex $\kappa$ bound to the full version of the paper. 
\begin{corollary}
\label{thm:main-kappa-complex}
Let $\mA \in \C^{d \times n}$  with $d \geq 4$ be a matrix with non-collinear columns, then $n \leq O( d^4 \kappa_{\mA}^2)$.
\end{corollary}

While \Cref{thm:real-rep} and \Cref{thm:complex-rep} require the use of
difficult matroid machinery, our proof of \Cref{thm:complex-rep-ours} relies
only on comparatively simpler geometric techniques. 

The main technical tool we require is an estimate on the maximum number of
distinct lines passing through some point of a $d$-dimensional complex point
set (the reader may restrict to real point sets for intuition, as the proofs
are identical in this case). This type of quantity has been intensively
studied in discrete geometry, especially for low dimensional point sets. 
We give precise definitions below:  

\begin{definition}
For point set $S \subseteq \CCC^d$, a line $L$ in $S$ is a maximal subset of $S$ having affine dimension $1$. Letting $\lines(S)$
denote the set of lines in $S$, we let $\maxlines(S) := \max_{s \in S} |\{L \in
\lines(S): s \in L\}|$ denote the maximum number of distinct lines in $S$
through any single point. We further define $f(d) \in [0,1]$ to be largest
number such that for any $S \subseteq \CCC^d$ with affine dimension $d$, we
have $\maxlines(S) \geq f(d)|S|$.     
\end{definition}

\begin{figure} \label{fig:LinesExample}
 \begin{center}
  \includegraphics{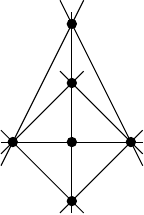} \hspace{2cm}
  \includegraphics{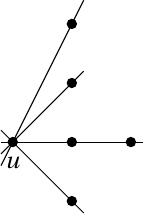}
\end{center}
  \caption{Left: all the lines defined by some points $S \subseteq \mathbb{R}^2$. Right: $\maxlines(S)=4$ attained by $u$.}
\end{figure}

The following simple lemma gives a direct connection between estimates on
$f(d)$ and the existence of minors with a large number of non-collinear
elements in complex representable matroids.

\begin{lemma}
\label{lem:find-minor}
Let $\cM$ be simple rank $d \geq 3$ complex representable affine matroid on $n$
elements. Then, for $2 \leq k \leq d-1$, $\cM$ contains a rank $k$ minor with
at least $n \prod_{r=k}^{d-1} f(r)$ non-collinear elements. In particular,
$\cM$ contains a line of length $n \prod_{r=2}^{d-1} f(r)$ as a minor.
\end{lemma}

The proof of the above lemma uses the affine representation of the matroid
$\cM$. That is, we use a point set $S = \{v_1,\dots,v_n\} \subseteq
\CCC^{d-1}$ such that a set $I \subseteq [n]$ is independent in $\cM$ if and only if
the points $\{v_i: i \in I\}$ are affinely independent. Note that the affine dimension of
$S$ will always be one less than the rank of $\cM$. From here, if $v_i$ is
the point with $\maxlines(S)$ distinct lines passing through it, one observes
that the rank $d-1$ minor $\cM / \{i\}$ of $\cM$ has exactly $\maxlines(S)
\geq f(d-1) n$ nonzero and non-collinear elements. To get the lower bound
for all ranks, one simply continues this process inductively. 

By the above lemma, if $\cM$ avoids a line of length $l$, then $n
\prod_{r=2}^{d-1} f(r) \leq l \Leftrightarrow n \leq l/\prod_{r=2}^{d-1}
f(r)$. Thus, lower bounds on $f(d)$ will allow us to deduce upper bounds for
\Cref{thm:complex-rep-ours}.    

As alluded to above, bounds on $\maxlines(S)$ have been intensively studied within the
field of discrete geometry (mainly for low dimensional point sets). In
particular, Dirac's conjecture states that for any \emph{real} point set $S
\subseteq \R^2$ of affine dimension $2$, that $\maxlines(S) \geq \lfloor
|S|/2 \rfloor+1$ for $|S|$ large enough. The best current bound due to
Han~\cite{han2017note}, based on the work of
Langer~\cite{langer2003logarithmic}, also applies to complex point sets
(see~\cite[Corollary 1.2]{de2018spanned}) and gives the following:

\begin{restatable}{theorem}{thmuno}\label{thm:low-dim}(\cite{han2017note})
For $S \subseteq \CCC^2$ of affine dimension $2$, $\maxlines(S) \geq \lfloor
|S|/3 \rfloor + 1$. In particular, $f(2) \geq 1/3$.
\end{restatable}

Note that the incidence structure of points and lines of a point set $S \subseteq
\CCC^d$ is preserved under a generic projection onto $\CCC^2$. Therefore, $f(d)
\geq f(2) \geq 1/3$ for $d \geq 2$.

Using this lower bound for all $d$, we would only be able to prove a bound of
$\prod_{r=2}^{d-1} f(r)^{-1} l \leq 3^{d-2} l$ for
\Cref{thm:complex-rep-ours}, which is much weaker than claimed. 

Fortunately, much better bounds for the high dimensional setting were given
by Dvir,  Saraf and Wigderson~\cite{dvir2014improved} (improving over the
preceeding work of Barak, Dvir, Widgerson and Yehudayoff~\cite{barak2011rankbounds}). The following
paraphrases their main result in the present language: 

\begin{theorem}[\cite{dvir2014improved}] \label{thm:dvir2014SGconfig}%\label{lemma: SG configuration lines lemma}
For a point set $S \subseteq \CCC^d$ with affine dimension $d$, we have
$\maxlines(S) \geq (1 - \frac{12}{d})|S|$. In particular, $f(d) \geq
1-\frac{12}{d}$. 
%    Thus $\maxlines(S) \geq (1 - \frac{12}{d})m$ for every full dimensional set $S \subset \mathbb{R}^d$. 
\end{theorem}

Combining \Cref{thm:low-dim} for low dimensions together with
\Cref{thm:dvir2014SGconfig} above for high dimensions, one can deduce a bound
of $O(d^{12} l)$ for \Cref{thm:complex-rep-ours}, falling short of our
claimed bound. This brings us to our final contribution which is an improved
bound for the setting considered by Dvir et al~\cite{dvir2014improved}.

Their bound is proved by studying the following family of sparse design
matrices:
\begin{definition}
  A matrix  $\mA \in \CCC^{m \times n}$ is a \emph{$(q,k,t)$-design matrix}  if
  (i) for all rows $i \in [m]$ one has $|\supp(\mA_{i,\bullet})| \leq q$;
  (ii) for all columns $j \in [n]$ one has $|\supp(\mA_{\bullet,j})| \geq k$ and
  (iii) for all column indices $j_1 \neq j_2$ one has $|\supp(\mA_{\bullet,j_1}) \cap \supp(\mA_{\bullet,j_2})| \leq t$. Here $\mA_{i,\bullet}$ denotes the $i^{th}$ row of $\mA$ and $\mA_{\bullet,j}$ denotes the $j^{th}$ column of $\mA$.
\end{definition}
Dvir, Saraf and Widgerson~\cite{dvir2014improved} proved that any such $(q,k,t)$-design matrix $\mA$ has
$\rank(\mA) \geq n - \frac{ntq(q-1)}{k}$. We improve the loss term by a factor of $q$ via a careful accounting of column versus row sizes.

\begin{theorem} \label{thm:RankOfDesignMatrix}
For any $(q,k,t)$-design matrix $\mA \in \CCC^{m \times n}$ one has $\rank(\mA) \geq \frac{n}{1+\frac{t(q-1)}{k}} > n - \frac{nt(q-1)}{k}$.
\end{theorem}
% This improves the bound of $\textrm{rank}(\mA) \geq n - \frac{ntq(q-1)}{k}$ proven by Dvir, Wigderson and Saraf~\cite{dvir2014improved}.

The arguments from \cite{barak2011rankbounds,dvir2014improved} reduce the
task of lower bounding $f(d)$ to lower bounding the rank of
$(3,3k,6)$-designs (see Lemma~\ref{lem:ExistenceOfAffineDepMatrixA}). The
improved rank lower bound given above then immediately yields the desired bound on
$f(d)$. 

\begin{theorem}\label{thm:SGConfigLinesLemmaImproved}
For a point set $S \subseteq \CCC^d$ with affine dimension $d$, we have
$\maxlines(S) \geq (1 - \frac{4}{d+1})|S|$. In particular, $f(d) \geq
1-\frac{4}{d+1}$. 
\end{theorem}

Using this improved theorem, we are finally able to deduce the claimed bound
for \Cref{thm:complex-rep-ours}. 

\subsection{Further Applications}

Our improved lower bound on $f(d)$ allows us to recover a slight quantitative
improvement on a result of~\cite{dvir2014improved} on the number of ordinary
flats in a point configuration. Working in matroid language, we recall that a
$k$-flat $F \subseteq \mathcal{M}$ is a set satisfying $\rank(F)=k$ and that
contains all the elements in $\mathcal{M}$ that are dependent with it (think of
the affine span of a set of points in a real or complex representation).
Equivalently, $\rank(F \cup \{v\}) = k+1$ for all $v \in \mathcal{M} - F$. $F$
is an \textit{ordinary} flat if it can be expressed as $F = H \cup \{v\}$,
where $H$ is a $k-1$ flat and $v$ is an element in $\mathcal{M}$. In this
case, we say that $F$ is an ordinary flat passing through $H$. 

Using a similar inductive strategy as in \Cref{lem:find-minor}, we get the
following lower bound on the number of ordinary flats passing through some
fixed flat. 

%     A flat of the set, $flat(e_1, \hdots, e_k) \cup \{v\}$, is called \textit{ordinary} in $\mathcal{M}$ if and only if $v$ is not parallel with any element in $\mathcal{M'}$ the matroid $\mathcal{M}$ contracted on $\{e_1, \hdots, e_k\}$.
% 

\begin{restatable}{corollary}{cortwo}\label{cor: inductiveClaim}
Let $\mathcal{M}$ be a simple rank $d$ complex representable matroid on $n$
elements. Then, for any $\epsilon \in [0,1]$ and $k \in [d]$ satisfying
$2\prod_{i=1}^{k-1} f(d-i)-1 \geq \epsilon$, there exists a $(k-1)$-flat $H
\subseteq \mathcal{M}$ with at least $\epsilon n$ ordinary $k$-flats passing
through it. In particular, for $d \geq 9$, any choice $1 \leq k \leq
(1-\epsilon)(d-3)/8$ satisfies the requirement.
\end{restatable}

\Cref{cor: inductiveClaim} is a lower bound on the number of ordinary flats in
a rank $d$ point set. This is essentially the contrapositive of what is shown in
\cite[Theorem 1.14]{dvir2014improved}, where they upper bound the dimension of a
point set given the number of such ordinary flats.  

As another application of the column bound for $\kappa_A$ bounded matrices, we show the following relation to integer programming proximity bounds. A common tool for algorithms and complexity analysis of integer programs is the \emph{Graver Basis} of the constraint matrix $A$. The Graver basis of  $A$ denoted $\mathcal{G}(A)$ is the set of $⊑$-minimal integer kernel elements in $\{x ∈ \ZZ^n : Ax = 0 , x \neq 0\}$ \cite{graver1975foundations}. The partial order $⊑$ is defined on $\ZZ^n$ such that for any $u, v \in \ZZ^n$, $u ⊑ v$ if $|u_i| ≤ |v_i|$ and $u_iv_i ≥ 0$ for each $i \in [n]$. We show that our column bound gives a proximity bound that depends only on the number of rows and Graver Basis of the constraint matrix. Proximity bounds and Graver Basis size bounds have historically had similar proof strategies. The following corollary indicates that there is indeed a polynomial relation between proximity and the Graver Basis independent of the number of columns of $A$.
\begin{restatable}{corollary}{corone}\label{cor: graverBasisCor}
Let an integer program be defined by a description $$IP = \min \{c^\T x: Ax = b, x \in [0, u], x \in \mathbb{Z}^n\}$$
    for some integer matrix $A \in \ZZ^{d \times n}$, $b, u \in \ZZ^n$. Let $x^{LP}$ be the optimal solution to the LP relaxation and $x^{IP}$ be the closest feasible integral solution, assuming that one exists. Let $\mathcal{G}(A)$ be the set of Graver basis vectors of $A$ and $g_\infty(A) = \max\{\Vert g \Vert_\infty: g \in \mathcal{G}(A)\}$ be the maximum entry of any Graver basis element. Then $$\Vert x^{IP} - x^{LP}\Vert_\infty \leq O(d^4g_\infty(A)^4).$$ 
\end{restatable}
 In particular, this corollary follows from a bound on the number of columns of $A$ by $O(d^4  g_\infty(A)^3)$. Finally, one may note that diameter bounds for polyhedra which depend on the number of columns can also be replaced by a bound that depends only on the circuit imbalance and the number of constraints of the LP matrix using our column number bound.

\subsection{Open Questions}
In terms of open questions, many still remain. One natural question is whether we can find other excluded minors in terms of the imbalance measure $\kappa_\mA$ in order to get closer to the right bound on the number of elements of a $\kappa$-bounded matroid. Several minor exclusion properties were shown in \cite{paat2024column} for $\Delta$-modular matrices. Some of these minors do not apply directly to $\kappa$-bounded matroids as $\kappa$-boundedness is closed under direct summation. 
Interestingly, $\kappa$ is also a self-dual measure, such that the $\kappa$ measure of any subspace is equal to that of its orthogonal complement. This means that any matroid is $\kappa$ bounded if and only if the dual matroid is $\kappa$ bounded such that matroid duality can be easily exploited. The class of $\Delta$-modular matroids was also shown to be closed under duality \cite{oxley20222}. 

An obvious direction of continuation is that of strengthening the bound on the number of spanned lines intersecting at a single point for any point configuration in $\C^d$. In particular, improved lower bounds on this quantity even by a constant factor would imply polynomial improvement in the column bounds on $\kappa$-bounded matroids and $\Delta$-modular matrices. A clear first step would be to extend the current bounds to include what are called \say{special} lines of the point configuration, or lines that include 3 or more points. Such lines are excluded from the analysis in this work.

\section{Matroid Theory Preliminaries}
\label{sec:matroid-prelims}

A \emph{matroid} is a pair $\mathcal{M} = (S, \mathcal{I})$ such that $\{ \emptyset \} \subseteq \mathcal{I} \subseteq 2^{S}$ and 
\begin{enumerate}
    \item If $I \in \mathcal{I}$ and $J \subseteq I$ then $J \in \mathcal{I}$ and 
    \item If $I, J \in \mathcal{I}$ and $|J| > |I|$, then there exists an element $e ∈ J \setminus I$ such that $I \cup \{e\} \in \mathcal{I}$.
\end{enumerate}

A set $I \in \mathcal{I}$ is called an \emph{independent set} of the matroid and a set $I \subseteq S$ with $I \notin \mathcal{I}$ is called \emph{dependent}. A minimal inclusion wise dependent set is called a \emph{circuit}. A matroid can be seen as an abstraction of a matrix over a field. 

We recall that a matroid is \emph{representable} over a field $\mathbb{F}$
if there exists a matrix $\mA = (a_1,\dots,a_n) \in \mathbb{F}^{d
\times n}$ such that the subsets $S \subseteq [n]$ for which the
columns $\{a_i: i \in S\}$ are linearly independent are in bijection with the independent sets in $\mathcal{I}$. The definition allows for two elements to be represented by vectors that are scalar multiples of each other. Such pairs of elements are called \emph{parallel} in $\mathcal{M}$. The $\vec{0}$ vector, or a dependent set containing only one element, is called a \emph{loop}. We call a matroid \emph{simple} if it has no loops and no parallel elements.
Recall that vectors $\{a_1, \hdots, a_l\} \subset \mathbb{F}^d$, are called \emph{affinely dependent} if there exist some $\{\lambda_1, \hdots, \lambda_l\} \subset \mathbb{F}$ such that $\sum_i \lambda_i a_i = 0$ and $\sum_i \lambda_i = 1$. The \emph{affine independence matroid} is defined identically to representable matroids with respect to affine independence rather than linear independence. The affine dimension of a set of vectors is the maximum number of affinely independent vectors in the set. Note that the affine dimension is 1 more than the linear dimension of the set.

Throughout this paper, we will use a few matroid operations. Specifically, a \emph{contraction} of $\mathcal{M}$ by a non-loop element $e$ is given by the matroid $\mathcal{M}/e := (S\setminus e, \mathcal{I'})$ where for $I \subseteq S$ one has $I  \in \mathcal{I'}$ if and only if $I \cup \{e\} \in \mathcal{I}$. A \emph{simplification} of a matroid $\mathcal{M}$ denoted by $\si(\mathcal{M})$ refers to restricting $S$ to a maximal subset of elements $S' \subseteq S$ such that the set of independent sets over $S'$ contains no loops or parallel elements. We also denote $\mathcal{E}(\mathcal{M})$ as the size of the simplification of $\mathcal{M}$. A \emph{minor} of a matroid $\mathcal{M}$ is another matroid $\mathcal{N}$ that is obtained from $\mathcal{M}$ by a sequence of deletion and contraction operations. Lastly, we denote $U_{k, q}$ by the \emph{uniform matroid} on $q$ elements of rank $k$. That is, every subset of size $k$ is an independent set in $U_{k, q}$.

Representable matroids can equivalently be defined via affinely independent sets of the matrix representation. This is shown by the following lemma. 

% Perhaps there is some confusion regarding what the precise relationship between matroids and systems of affine flats, the latter being what we have mostly used in the remainder of the paper. The relationship turns out to be simple.

% Firstly, a standard definition is that of an affine matroid. An affine
% matroid $\cM$ over a field $\F$ is defined by points $(v_1,\dots,v_n) \in
% \F^d$, where the independent sets $\cI$ are precisely the affinely
% independent subsets of $(v_1,\dots,v_n)$. 

\begin{lemma}\label{lem:aff-matroid}
For a field $\F$ and a matroid $\cM$ on $n < |\F|$ elements, $\cM$ is an affine
matroid over $\F$ if and only if $\cM$ is a finite loopless linear matroid
over $\F$.
\end{lemma}    
\begin{proof}
If $\cM$ is an affine matroid, then one can transform it into a linear
matroid by appending a row of all ones to its representation. For the reverse
direction, any loopless $n$ linear matroid over $\F$ is represented by a
matrix $\mA = (a_1,\dots,a_n) \in \F^{d \times n}$ with no zero column. Given
such a representation, using that $n < |\F|$, one can construct a vector $y
\in \F^d$ such that $r := y^\T \mA$ has only nonzero entries.
Replacing $\mA$ by $\mA{\rm diag}(r^{-1})$ where $r^{-1}$ if the component wise inverse of $r$ (note that this does not change
the underlying linear matroid), it is easy to check that $\mA$ is row
equivalent to a matrix containing a row of all ones. Thus, the linear matroid
induced by $\mA$ has an affine representation.   
\end{proof}

Note that when the underlying field is $\R$ or $\C$, then the condition $n <
|\F|$ is trivially satisfied since these fields are infinite. Thanks to the
equivalence given in Lemma \ref{lem:aff-matroid}, we will routinely switch
between affine and linear representations of matroids whenever convenient.

\section{From many points on a line to the \texorpdfstring{$\kappa_{\mA}$}{kappa} measure}

We now show the connection between the condition measure $\kappa$ and the set of lines defined by the point set $S$. We will let $S \subset \mathbb{C}^{d-1}$ be a full dimensional set and $\mathcal{M}$ be the affine matroid on $S$ (of rank $d$). We will show the relationship to the condition measure of the matrix with columns $S$ by first arguing that when $S$ is large, there is a large rank 2 minor of $\mathcal{M}$.

\subsection{Large rank 2 minor}\label{sec:large-rank-2-minor}
 We let $S \subset \mathbb{C}^{d-1}$ be the ground set of the affine matroid $\mathcal{M}$ such that $|S| = n$. We claim that there exists an element $e \in S$ such that the simplification of the contracted matroid $\mathcal{M} / e$ has at least $f(d-1) \cdot n$ elements. Let $\mA$ be the matrix with column vectors $S$. As a reminder, we denote $\textrm{maxlines}(S)$ as the maximum number of distinct lines defined by $S$ through any single point in $S$. 

We have that $\maxlines(S) \geq f(d-1) \cdot n$ by definition. Let $e \in S$ be a point such that the number of distinct lines incident to $e$ in $S$ is $\maxlines(S)$. Then for each $v_i, v_j \in S$ such that $v_i, v_j$ lie on different lines through $e$, the set $\{v_i, v_j\}$ is independent in $\mathcal{M}/e$. Thus a set of distinct lines through $e$ gives a set of pairwise non-parallel elements in $\mathcal{M}/e$. As $\mathcal{M}$ was a simple matroid, there are no loops in $\mathcal{M}/e$ so $\mathcal{E}(\mathcal{M}/e) \geq \maxlines(S)$. 
 
 In order to give a bound on $|S|$ in terms of $\kappa_{\mA}$, we will need the following lemma.

\begin{lemma}\label{lem:matroidHyperplaneArrangements}
  For a simple matroid $\mathcal{M} = (S, \mathcal{I})$ where $|S| = n$, $\rank(\mathcal{M}) = d$, and $\mathcal{M}$ is complex representable, there exists a simple matroid $\mathcal{M'} = (S', \mathcal{I}')$ obtained via a sequence of contractions and simplifications on $\mathcal{M}$ such that $$|S'| \geq \left(\displaystyle\prod_{i = 2}^{d-1} f(i) \right)\cdot n$$
  and $\rank(\mathcal{M'}) = 2$. 
\end{lemma}

\begin{proof} 
    We prove the lemma by induction on the rank $d$ of the matroid. As a base case, let $d = 3$ and note that $S$ itself trivially satisfies the claim as $f(i) \in [0,1]$. 
    
    For the inductive step, we assume that the lemma holds for $k = d-1$. Next, we consider a matroid $\mathcal{M}$ of rank $d$. Note that by the assumption $\maxlines(S) \geq f(d-1) \cdot n$, there exists an element $e$ such that $\mathcal{E}\left(\mathcal{M}/e\right) \geq f(d-1) \cdot n$ where $\mathcal{E}(\mathcal{N})$ refers to the size of the ground set of $\si(\mathcal{N})$.

    Note that by contraction on $e$ and as $\mathcal{M}$ was simple, $\rank(\si(\mathcal{M}/e)) =d-1$. By the induction hypothesis, for $\mathcal{M'} := \si(\mathcal{M}/e)$ there exists a sequence of contractions and simplifications to obtain a matroid $\mathcal{M''} = (S'', \mathcal{I}'')$ of rank $2$ and size $|S''| \geq \big(\prod_{i = 2}^{d-2} f(i) \big)\cdot |S'| = \big(\prod_{i = 2}^{d-1} f(i) \big)\cdot n$. As $\mathcal{M'}$ itself was obtained via a contraction and restriction on $\mathcal{M}$, we conclude the lemma.  
\end{proof}

Note also that Lemma \ref{lem:find-minor} follows from the above proof by truncating the induction at $d = k+1$ and taking the product $\big(\prod_{i = k}^{d-1} f(i) \big)\cdot n$. 

\subsection{Many distinct flats}\label{sec:many-distinct-flats}

From a geometric perspective, section \ref{sec:large-rank-2-minor} shows that a full dimensional point set $S \subset \mathbb{R}^d$ of size $n$ contains many hyperplanes defined by $S$ that pass through a $(d-2)$-dimensional flat. Recall that a \emph{flat} is an affine subspace, not necessarily going through the origin.

\begin{figure}
 \begin{center}
  \includegraphics{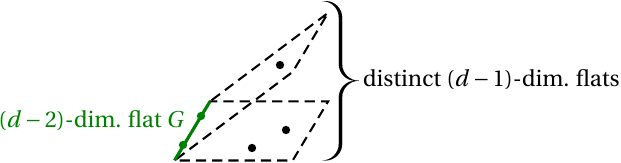} 
\end{center}
  \caption{Visualization of many $(d-1)$-dimensional flats containing $G$ for $d=3$.}\label{fig:ManyFlats}
\end{figure}

%This can be seen by noting that the inductive step finds a vector $v^\ast$ such that $v^\ast$ is in at least $f(d) \cdot n$ many distinct lines defined by $S$. The vectors are then centrally projected with center $v^\ast$, such that the unique set of projected points form the inductive hypothesis in one dimension lower. Inductively, this results in many $(d-1)$-dimensional hyperplanes determined by $S$ that contain a $(d-2)$-dimensional affine space. A homogenizing step then gives the same claim in terms of subspaces. 
This can be phrased as the following geometric corollary of Lemma \ref{lem:matroidHyperplaneArrangements}. See also figure \ref{fig:ManyFlats}.

\begin{corollary}\label{lem:hyperplaneArrangements}
  For $S \subseteq \mathbb{R}^d$ such that $\spa(S) = \RR^d$, $|S| = n$, there exists a $(d-2)$-dimensional affine subspace $G$
  so that  at least $$\left(\displaystyle\prod_{i = 2}^{d-1} f(i) \right)\cdot n$$ distinct $(d-1)$-dimensional hyperplanes determined by $S$ contain $G$. 
\end{corollary}

\subsection{From a large rank 2 minor to a \texorpdfstring{$\kappa_{\mA}$}{kappa} bound}

Using Lemma \ref{lem:matroidHyperplaneArrangements}, namely that there exists a large rank 2 minor of $\mathcal{M}$, we can now give a bound on the number of columns of $\mA$. To do this, we will need to use two lemmas that show $\kappa_\mA$ only decreases under contraction and simplification, and that the resulting rank 2 matroid has a bound in terms of $\kappa_\mA$. In particular, we prove Lemma \ref{lem:kappa-minor} and show that matroids represented by a real matrix (respectively complex matrix) with circuit imbalance measure at most $\kappa_\mA$ form a minor closed family that excludes a line of length $O(\kappa_\mA)$ (respectively $O(\kappa_\mA^2)$).

As before, we let $\mathcal{M} = (S, \mathcal{I})$ be the affine matroid with ground set $S$ and complex matrix representation $\mA$. As a reminder, this means that the affinely independent subsets of columns of $\mA$ are in bijecion with the independent sets in $\mathcal{M}$. From now on, to get the bounds in terms of $\kappa_\mA$, we will assume that $\mathcal{M}$ is a complex representable matroid. By Lemma \ref{lem:matroidHyperplaneArrangements}, there exists a matroid $\mathcal{M'} = (S', \mathcal{I'})$ of rank $2$ obtained via a sequence of contractions and restrictions on $\mathcal{M}$ where $|S'| \geq \left(\prod_{i = 2}^{d-1} f(i) \right) \cdot n$. 

We denote the complex representation of $\mathcal{M'}$ by $\mA'$. 
We now argue that the circuit imbalance measure only decreases from $\ker(\mA)$ to $\ker(\mA')$. For this, we use the results of \cite{ekbatani2022circuit} translated to linear subspaces. 

    \begin{lemma}[\protect{Paraphrased from \cite[Proposition 3.17]{ekbatani2022circuit}}]\label{lem:kappa-on-subspaces}
        For any linear subspace $W \subseteq \RR^n, J \subseteq [n]$ we have that $\kappa_{\pi_J(W)} \leq \kappa_W$, where $\pi_J(W)$ is the projection of $W$ onto the coordinates $J$.
    \end{lemma}
    Note that when referring to a matrix $\mA$, $\kappa_{\mA} := \kappa_{\ker(\mA)}$. We thus show that there exists $J \subseteq [n]$ such that $\ker(\mA') = \pi_J(\ker(\mA))$ and conclude that $\kappa_{\mA'} \leq \kappa_{\mA}$ by Lemma \ref{lem:kappa-on-subspaces}.

    We first assume that $\mathcal{M'}$ is obtained by a single contraction and simplication of $\mathcal{M}$, that is $\mathcal{M'} = \si(\mathcal{M} / e)$. The matroid $\mathcal{M'}$ can then be represented by $\mA'$, a matrix obtained from $\mA$ by removing the column corresponding to $e$, the first row, and any resulting collinearities. Note that this assumes via non-singular transformation that $e$ corresponds to $e_1$ the canonical unit vector. This gives a representation $\mA'$ of the matroid $\mathcal{M'}$ (see e.g. \cite{gordon2012matroids}, Chapter 6).

    Let $\mA = \left( a_1, \hdots, a_n \right)$ where here $a_i$ is a column of $\mA$. Let $J$ be an index set such that for every $i \in J$, $v_i \in \mathcal{M}$ if and only if $ v_i \in \mathcal{M'}$. Then for any minimal affine dependency $g \in \ker(\mA)$, we have that $g_J \in \ker(\mA')$. Moreover, for any vector $h \in \ker(\mA')$, we can extend $h$ with zeros until it has $n$ components and let $h_n = - \left[\sum_i h_i a_i\right]_1$. We thus get a kernel element $h \in \ker(\mA)$ with $h_J \in \ker(\mA')$. This shows that $\ker(\mA') = \pi_J(\ker(\mA))$.

    By Lemma \ref{lem:kappa-on-subspaces}, the value of $\kappa_{\mA}$ can only be larger than $\kappa_{\mA'}$ for $\mathcal{M'}$ obtained by a single contraction and simplification. The claim then extends to multiple iterations of contraction and simplification by a repetition of the argument.

    We note that $\mathcal{M'} = (S', \mathcal{I}')$ is a rank $2$ complex representable matroid. We will use the following bound on $\kappa$ to bound the number of elements of $S'$. 

    \begin{lemma}\label{lem:kappa-bound-2d}
    Let $\mA \in \R^{2 \times l}$ be a matrix with nonzero and non-collinear columns. Then $\kappa_{\mA} \geq \frac{l}{4\pi}$.
    \end{lemma}
    \begin{proof}
    We denote by $\mA_{i, j}$ the matrix $\mA$ restricted to columns $i$ and $j$. We denote by $\theta(v, w)$ the euclidean angle between vectors $v,w \in \RR^2$. When not specified, $\Vert \cdot \Vert$ refers to the $l_2$ norm. By a reordering of the columns of $\mA$, we may assume that $\det(\mA_{1, 2})$ is the maximum $2 \times 2$ subdeterminant of $\mA$. Next note that a nonsingular transformation on $\mA$ does not change the linear space $\ker(\mA)$ so $\kappa_{\mA} = \kappa_{\mA_{1, 2}^{-1} \cdot \mA}$. Consider the columns of $\mA_{1, 2}^{-1} \cdot \mA $, which are a set of $l$-many non-collinear and nonzero columns. Moreover, the set of vectors all have entries bounded in absolute value by $1$ by the choice of $\det(\mA_{1, 2})$ as maximal. As the pairwise angle between any two columns in this set is strictly positive, and by the pigeonhole principle, there must exist columns $a_i, a_j$ for some $1 \leq i < j\leq l$ such that the angle between them is at most $\frac{\pi}{l}$. 
    Finally, define two vectors $g_1, g_2 \in \RR^2$ by the columns on the matrix $(\mA_{1, 2}^{-1} \cdot \mA )_{i, j}^{-1}$. The entries of $(g_1, g_2)$ are thus made up of the entries of $(a_i, a_j)$ multiplied by $\frac{1}{|\det(\mA_{1, 2}^{-1} \cdot \mA )_{i, j}|}$. Without loss of generality

    \begin{align*}
    |g_{11}|  &\geq \frac{\Vert a_j\Vert /2}{|\det(\mA_{1, 2}^{-1} \cdot \mA )_{i, j}|} \\
    & = \frac{\Vert a_j\Vert /2}{\Vert a_i\Vert \Vert a_j\Vert |\sin(\theta(a_i, a_j))|} \\
    & \geq \frac{l}{2\pi \Vert a_i \Vert} \\
    & \geq \frac{l}{4 \pi}
    \end{align*}
    where the last inequality comes from the entries of $\mA_{1, 2}^{-1}\cdot \mA$ being bounded by $1$ in absolute value and where we have used $\sin(x) \leq x$. 
    Taking the circuit $-1 \cdot a_1 + g_{11} \cdot a_i + g_{12}\cdot a_j = \vec{0}$ gives a circuit with maximum entry $\frac{l}{4\pi}$ and minimum entry at most $1$.
    \end{proof}

    Then for $\mathcal{M'}$ represented by $\mA'$, and by Lemmas \ref{lem:matroidHyperplaneArrangements}, \ref{lem:kappa-on-subspaces}, and \ref{lem:kappa-bound-2d}, $\kappa_{\mA'} \geq \frac{|S'|}{4\pi}$ such that $\kappa_{\mA} \geq \frac{|S'|}{4\pi} \geq \frac{1}{4\pi} \left(\prod_{i = 2}^{d-1} f(i) \right) \cdot n$. As $|S| = n$, we get a bound on the size of $S$ of $|S| \leq 4\pi \cdot\kappa_{\mA} \left(\prod_{i = 2}^{d-1} \frac{1}{f(i)}\right) $.

    For a complex matrix representation $\mA$, we deduce Corollary \ref{thm:main-kappa-complex} via another determinant bound. 
\begin{proof}[Proof of Corollary \ref{thm:main-kappa-complex}]
    We upper bound $|\det( \mA _{i, j})|$ for some choice of indices $1 \leq i < j\leq l$ and conclude as in the real matrix proof of Lemma \ref{lem:kappa-bound-2d}. 

    First, note that as in the real valued case by nonsingular transformation, we may assume that all entries of $\mA$ have modulus less than or equal to $1$. As $\mA$ is a complex matrix, let $\mA_j = \left(\begin{matrix}
        r_{1j}e^{i\theta_{1j}} \\ r_{2j}e^{i\theta_{2j}}
    \end{matrix}\right)$ be the $j^{th}$ column of $\mA$. Define the matrix $$\tilde{\mA} := \mA \cdot \left(\begin{matrix}
        e^{-i\theta_{11}} & 0 & \hdots & 0 \\
        0 & e^{-i\theta_{12}} & \hdots & 0 \\
        \vdots & \vdots & \ddots& \vdots \\
        0 & 0 & \hdots & e^{-i \theta_{1l}}
    \end{matrix}\right)$$ and note that for every pair of indices $i, j$, $|\det(\mA_{i, j})| = |\det(\tilde{\mA}_{i, j})|$.

    Next, note that for each index $j \in [l]$, $\tilde{\mA}_j = \left(\begin{matrix}
        r_{1j} \\ r_{2j}e^{i(\theta_{2j} - \theta_{1j})}
    \end{matrix}\right)$. Define the angle $\tilde{\theta}_j : = (\theta_{2j} - \theta_{1j})$ for every $j$. Note that $\{\tilde{\theta}_1, \hdots, \tilde{\theta}_l\}$ is a set of $l$ arguments all within the interval $[0, 2\pi]$. By the pigeonhole principle, there exists $\sqrt{l}$ many arguments $\{\tilde{\theta}_{k_1}, \hdots, \tilde{\theta}_{k_{\sqrt{l}}}\}$ with absolute pairwise differences less than $\frac{2\pi}{\sqrt{l}}$. 

    Consider the set of vectors $\{R_{k_1}, \hdots, R_{k_{\sqrt{l}}}\}$ defined by $R_{k_i} := \left(\begin{matrix}
        r_{1k_i} \\ r_{2k_i}
    \end{matrix}\right)$, that is the vector of moduli of $\tilde{\mA}_{k_i}$. As this forms a set of real vectors, and by the pigeonhole principle, there exists a pair of vectors in the set with $|\det(R_k, R_j)|\leq \frac{2\pi}{\sqrt{l}} \Vert R_j \Vert_2^2$. 

    Finally, we bound $|\det(\tilde{\mA}_{k, j})|$ as follows:
    \begin{align*}
        |\det(\tilde{\mA}_{k, j})| & = \left|\det\left(\left( \begin{matrix}
            1 & 0 \\
            0 & e^{-i \tilde{\theta}_k}
        \end{matrix}\right)\tilde{\mA}_{k, j}\right)\right| \\
        & = |\det\left(\begin{matrix}r_{1k} & r_{1j} \\
        r_{2k} & r_{2j}e^{i (\tilde{\theta}_j - \tilde{\theta}_k)}\end{matrix}\right)| \\
        & = \sqrt{r_{1k}^2r_{2j}^2 + r_{1j}^2r_{2k}^2 - 2 r_{1j}r_{1k}r_{2j}r_{2k}\cos(\tilde{\theta}_j - \tilde{\theta}_k)}.
    \end{align*}
    Using the approximation $\cos(x) \approx 1 - \frac{x^2}{2}$, we get that $|\det(\tilde{\mA}_{k, j})| \leq \sqrt{\det(R_j, R_k)^2 + r_{1j}r_{1k}r_{2j}r_{2k}(\tilde{\theta}_j - \tilde{\theta}_k)^2} \leq \sqrt{\det(R_j, R_k)^2 + \Vert R_j \Vert_2^4 (\tilde{\theta}_j - \tilde{\theta}_k)^2}$. By the upper bounds on the determinant and the difference in angles, we get that $|\det(\tilde{\mA}_{k, j})| \leq \frac{4\pi}{\sqrt{l}}\Vert R_j \Vert_2^2$.
    % As both the determinant and the difference in angles are upper bounded by $\frac{2\pi}{\sqrt{l}}$, we get the determinant bound $|\det(\tilde{\mA}_{k, j})| \leq \frac{4\pi}{\sqrt{l}}$. 
    Plugging this determinant bound into the analysis gives a circuit with maximum entry of the order $\sqrt{l}$ such that the complex column bound follows.

\end{proof}

\section{Proof of Theorem~\ref{thm:main-kappa}}

Next, we want to prove  Theorem~\ref{thm:complex-rep-ours} and show that for any $U_{2, l}$ minor free affine matroid on $n$ elements, one has $n \leq \frac{1}{4}d^4 l$.

To conclude Theorem \ref{thm:complex-rep-ours}, we need a bound on the function $f(d)$. Here we are assuming the bound of $f(d) \geq 1-\frac{4}{d+1}$ (which we still need to prove). There is the obvious issue with this bound that it is vacuous for $d \leq 4$. As a reminder, for small $d$ one can obtain a better bound via the following theorem:

\thmuno*

We can now conclude:
\begin{proof}[Proof of Theorem~\ref{thm:complex-rep-ours}]
    Let $\mathcal{M} = (S, \mathcal{I})$ be the affine matroid with ground set $S$ and representation $\mA$. By Lemma \ref{lem:matroidHyperplaneArrangements}, there exists a matroid $\mathcal{M'} = (S', \mathcal{I'})$ of rank $2$ obtained via a sequence of contractions and restrictions on $\mathcal{M}$ where $|S'| \geq \left(\prod_{i = 2}^{d-1} f(i) \right) \cdot n$. Thus if $\mathcal{M}$ excludes a rank 2 uniform minor of size $l$, then $n \cdot \left(\prod_{i = 2}^{d-1} f(i) \right)  \leq |S'| \leq l$ such that $n \leq \left(\prod_{i = 2}^{d-1} f(i) \right)^{-1}\cdot l$.

Next, we can use the bound of Theorem~\ref{thm:SGConfigLinesLemmaImproved} to get $f(i) \geq 1-\frac{4}{i+1}$.
For $2 \leq i \leq 4$ we use $f(i) \geq 1/3$ from \Cref{thm:low-dim}. Then for $d \geq 9$, we have that 
\[
n \leq l \cdot \prod_{i=2}^{d-1} \frac{1}{f(i)} \leq l\cdot (\prod_{i=2}^4 3)(\prod_{i=5}^{d-1} \frac{i+1}{i-3}) = l\cdot \frac{9}{40} d(d-1)(d-2)(d-3) \leq \frac{1}{4} ld^4.
\]
Using the same formula as above, one can verify numerically that the right hand
side is at most $l d^4/4$ for $d \in \{3,4,5,6,7,8\}$.
\end{proof}

Finally, we conclude Theorem \ref{thm:main-kappa} and show that for any full dimensional matrix $\mA \in \mathbb{R}^{d\times n}$, one has $n \leq \pi d^4 \kappa_\mA$. We have shown that $n \leq 4\pi \cdot \kappa_{\mA} \left(\prod_{i = 2}^{d-1} \frac{1}{f(i)}\right) $, and the above bound on $\left(\prod_{i = 2}^{d-1} f(i) \right)^{-1}$ concludes the Theorem.

\section{An improved rank lower bound for design matrices}

%We define a certain family of sparse matrices: 
% \begin{definition}
%   A matrix  $A \in \RR^{m \times n}$ is a \emph{$(q,k,t)$-design matrix}  if
%   (i) for all rows $i \in [m]$ one has $|\textrm{supp}(A_i)| \leq q$;
%   (ii) for all columns $j \in [n]$ one has $|\textrm{supp}(A^j)| \geq k$ and
%   (iii) for all column indices $j_1 \neq j_2$ one has $|\textrm{supp}(A^{j_1}) \cap \textrm{supp}(A^{j_2})| \leq t$.
% \end{definition}

In this section we will prove Theorem~\ref{thm:RankOfDesignMatrix} %the following:
and show that any $(q,k,t)$-design matrix $\mA \in \RR^{m \times n}$ one has $\rank(\mA) \geq n - \frac{nt(q-1)}{k}$.
%\begin{theorem} \label{thm:RankOfDesignMatrix}
%For any $(q,k,t)$-design matrix $A \in \RR^{m \times n}$ one has $\textrm{rank}(A) \geq n - \frac{nt(q-1)}{k}$.
%\end{theorem}
As mentioned earlier, this improves the bound of $\rank(\mA) \geq n - \frac{ntq(q-1)}{k}$ proven by Dvir, Wigderson and Saraf~\cite{dvir2014improved}.
We follow the same strategy as \cite{dvir2014improved} but do a more careful accounting of rows vs. columns which leads to the improvement. First note that a design matrix is only specified by its \emph{support} while it appears that we have no control over the
actual values $\mA_{ij}$.

\subsection{Matrix scaling}

Given a matrix $\mA \in \CCC^{m \times n}$ and vectors $\rho \in \RR_{> 0}^m$ and $\gamma \in \RR_{> 0}^{n}$, we define
a scaled matrix $\mB := SC(\mA,\rho,\gamma) \in \CC^{m \times n}$ whose entries are $\mB_{i,j} := \mA_{i,j} \cdot \rho_i \cdot \gamma_j$.
In other words, we obtain $\mB$ by multiplying all rows $i$ by $\rho_i$ and all columns $j$ by $\gamma_j$. It was proven by
Rothblum and Schneider~\cite{rothblum1989scalings} that a non-negative matrix $\mA$ can be rescaled to have specified row sums and
column sums if and only if a certain transportation problem is feasible. Our matrix is not non-negative but one can simply apply
\cite{rothblum1989scalings} to the matrix $\mA'$ with entries $\mA_{i,j}' := |\mA_{i,j}|^2$. We state the results of \cite{rothblum1989scalings}
translated to the real matrix with squared entries:

\begin{definition}
Let $\mA \in \CCC^{m \times n}$ and $r \in \RR_{\geq 0}^m$ and $c \in \RR_{\geq 0}^n$.
We say that $\mA$ is \emph{asymptotically $(r,c)$-scalable} if for any $\varepsilon>0$ there are vectors $\rho \in \RR_{> 0}^m$ and $\gamma \in \RR_{> 0}^{n}$ so that the rescaled matrix $\mB := SC(\mA,\rho,\gamma)$  satisfies
  \[
   |\|\mB_{i,\bullet}\|_2^2-r_i| \leq \varepsilon \quad \forall i \in [m] \quad \textrm{and} \quad |\|\mB_{\bullet,j}\|_2^2-c_j| \leq \varepsilon \quad \forall j \in [n]
  \]
\end{definition}

We also use the term \emph{asymptotically $(\leq r, c)$-scalable} if the
inequality $|\|\mB_{i,\bullet}\|_2^2-r_i| \leq \varepsilon$ is replaced by $\|\mB_{i,\bullet}\|_2^2
\leq r_i + \varepsilon$. While $\varepsilon > 0$ can be chosen as small as
desired, we note that its presence is in fact necessary. For example the
matrix
\[
\mA = \begin{pmatrix} 1 & 1 \\ 0 & 1 \end{pmatrix}
\]
is asymptotically $({\bf 1},{\bf 1})$-scalable, but not exactly (i.e. with $\varepsilon = 0$) $({\bf 1},{\bf 1})$-scalable.
For a matrix $\mA \in \CCC^{m \times n}$ we define the \emph{support graph} as the bipartite graph
$G = (U \sqcup V,E)$ with $U := [m]$ and $V := [n]$ and edges $E := \{ (i,j) : \mA_{i,j} \neq 0\}$.
\begin{theorem}[Paraphrased from \cite{rothblum1989scalings}] \label{thm:MatrixRescalingConditions}
  Let $\mA \in \CCC^{m \times n}$, $r \in \RR_{\geq 0}^m$ and $c \in \RR_{\geq 0}^n$. Then the following are equivalent
  \begin{enumerate}
  \item[(I)] $\mA$ is asymptotically $(r,c)$-scalable.
%  \item[(II)] $A$ is asymptotically $(\leq r,c)$-scalable.
  \item[(II)] The following transportation problem has a solution $z \in \RR_{\geq 0}^E$:
    \[
 \sum_{j: (i,j) \in E} z_{ij} = r_i \quad \forall i \in [m]; \quad \sum_{i: (i,j) \in E} z_{ij} = c_j \quad \forall j \in [n]
\]
%\item[(IV)] One has
%   \[
%    c(J) \leq r(N(J)) \quad \forall J \subseteq [n]
%   \]
%   where $c(J) := \sum_{j \in J} c_j$ and $N(J) := \{ i \in [m] \mid \exists j \in J: (i,j) \in E\}$ denotes the neighborhood of columns $J$ in the support graph.
   \end{enumerate}
\end{theorem}
% In the context of matchings $(IV)$ is usually called \emph{Hall's condition} and it is particularly easy to verify.

\begin{lemma} \label{lem:ScalingDesignMatrix}
A $(q,k,t)$-design matrix $\mA \in \CCC^{m\times n}$ is asymptotically $(\leq
\frac{q}{k} \mathbf{1}_m, \mathbf{1}_n)$-scalable.
\end{lemma}
\begin{proof}
Let $k_j = |\supp(\mA_{\bullet,j})| \geq k$, $j \in [n]$, denote the number
of nonzeros in the $j$th column. Examine the transportation solution $z_{ij}
= 1/k_j$, $\forall (i,j) \in \supp(\mA)$. By construction, the row and column sums satisfy
\begin{align*}
r_i &:= \sum_{j: (i,j) \in \supp(\mA)} z_{ij} \leq |\supp(\mA_{i,\bullet})|/k \leq q/k, i \in [m], \\
c_j &:= \sum_{i: (i,j) \in \supp(\mA)} z_{ij} = |\supp(\mA_{\bullet,j})|/k_j = 1, \forall j \in [n].
\end{align*}
By \Cref{thm:MatrixRescalingConditions}, we conclude $\mA$ is asymptotically $(r,{\bf 1}_n)$-scalable where $r \in [0,\frac{q}{k}]^m$. In
particular, $\mA$ is asymptotically $(\leq \frac{q}{k} {\bf 1}_m, {\bf 1}_n)$-scalable.   
\end{proof}

\subsection{Some linear algebra lemmas}

For a matrix $\mA \in \CCC^{m \times n}$, the \emph{Frobenius norm} is $\|\mA\|_F := (\sum_{i=1}^m \sum_{j=1}^n |\mA_{i,j}|^2)^{1/2}$.
The following is a well known argument to lower bound the rank of a matrix. The proof is standard; we give it here for
the sake of completeness:
\begin{lemma} \label{lem:RankVsTraceAndFrobNorm}
For a Hermitian positive semidefinite matrix $\mM \in \CCC^{n \times n}$ one has $\rank(\mM) \geq \frac{\Tr[\mM]^2}{\|\mM\|_F^2}$.
\end{lemma}
\begin{proof}
  Let $\lambda_1,\ldots,\lambda_r>0$ be all the positive eigenvalues of $\mM$. Then
  \[
  \Tr[\mM]^2 = \Big(\sum_{i=1}^r \lambda_i\Big)^2 \leq r \sum_{i=1}^r \lambda_i^2 = r \|\mM\|_F^2
  \]
  Rearranging for $r$ gives the claim.
\end{proof}

In order to effectively apply Lemma~\ref{lem:RankVsTraceAndFrobNorm} one would need to make sure that the off-diagonal
entries of the matrix $\mM$ are small. To do so we will use almost the same argument as \cite{dvir2014improved} but
deviate at the end and express the bound in terms of $\|\mA\|_F^2$ rather than the number $m$ of rows. 
\begin{lemma} \label{lem:OffdiagonalEntriesOfDesignMatrix}
  Let $\mA \in \CCC^{m \times n}$ be a $(q,k,t)$-design matrix with $\alpha := \max_{i=1,\ldots,m} \|\mA_{i,\bullet}\|_2^2$. Then $\mM := \mA^\ast\mA \in \CCC^{n \times n}$ satisfies
  \[
  \sum_{(j_1,j_2) \in [n]^2: j_1 \neq j_2} |\mM_{j_1,j_2}|^2 \leq t \Big(1-\frac{1}{q}\Big)\alpha \|\mA\|_F^2
  \]
\end{lemma}
\begin{proof}

 First fix two distinct indices $j_1,j_2 \in [n]$ and recall that $\mM_{j_1,j_2} = \left<\bar{\mA}_{\bullet,j_1},\mA_{\bullet,j_2}\right>$.
  We abbreviate the vector $a := (\bar{\mA}_{i,j_1}\cdot \mA_{i,j_2})_{i \in [m]}$.
  Then
  \[
  \left<\bar{\mA}_{\bullet,j_1},\mA_{\bullet,j_2}\right>^2 = \Big(\sum_{i=1}^m |a_i|\Big)^2 \leq \|a\|_1^2 \leq \underbrace{|\supp(a)|}_{\leq t} \cdot \|a\|_2^2 \leq t\sum_{i=1}^m |\mA_{i,j_1}|^2|\mA_{i,j_2}|^2. \quad (*)
\]
We use this insight to bound
\begin{eqnarray*}
  \sum_{(j_1,j_2) \in [n]^2: j_1 \neq j_2} |\mM_{j_1,j_2}|^2 &=& \sum_{(j_1,j_2) \in [n]^2: j_1 \neq j_2} \left<\bar{\mA}_{\bullet,j_1},\mA_{\bullet,j_2}\right>^2 \\
                                    &\stackrel{(*)}{\leq}& t \sum_{(j_1,j_2) \in [n]^2: j_1 \neq j_2} \sum_{i=1}^m |\mA_{i,j_1}|^2|\mA_{i,j_2}|^2 \\
                                    &=& t \sum_{i=1}^m \Big(\sum_{(j_1,j_2) \in [n]^2} |\mA_{i,j_2}|^2|\mA_{i,j_2}|^2- \sum_{j=1}^n |\mA_{i,j}|^4 \Big) \\
                                                         &=& t \Big( \sum_{i=1}^m \Big(\sum_{j=1}^n |\mA_{i,j}|^2\Big)^2 - \sum_{i=1}^m \sum_{j=1}^n |\mA_{i,j}|^4 \Big).\quad (**)
\end{eqnarray*}
As before for any $x \in \RR_{\geq 0}^n$ one has $\sum_{j=1}^n x_j^2 \geq \frac{1}{|\supp(x)|} (\sum_{j=1}^n x_j)^2$; 
we combine this fact and the assumption that all rows have support at most $q$ and may continue % together with $|\supp(\mA_i)| \leq q$.
\begin{eqnarray*}
 (**)                                   &\leq& t \Big(\sum_{i=1}^m \Big(\sum_{j=1}^n |\mA_{i,j}|^2\Big)^2 - \frac{1}{q} \sum_{i=1}^m \Big(\sum_{j=1}^n |\mA_{i,j}|^2\Big)^2 \Big) \\
                                        %&=& t \Big(1-\frac{1}{q}\Big) \sum_{i=1}^m \Big(\sum_{j=1}^n A_{ij}^2\Big)^2 \\
                                        &=& t\Big(1-\frac{1}{q}\Big) \sum_{i=1}^m \|\mA_{i,\bullet}\|_2^4 \\ &\leq& t\Big(1-\frac{1}{q}\Big) \alpha \sum_{i=1}^m \|\mA_{i,\bullet}\|_2^2 = t\Big(1-\frac{1}{q}\Big)\alpha \|\mA\|_F^2.
\end{eqnarray*}
This proves the claim.
\end{proof}

\subsection{The rank of design matrices}

Now we are ready to prove Theorem~\ref{thm:RankOfDesignMatrix}. 
\begin{proof}[Proof of \Cref{thm:RankOfDesignMatrix}]
  We fix a $(q,k,t)$-design matrix $\mA \in \CCC^{m \times n}$; our goal is to lower bound $\rank(\mA)$.
  Scaling $\mA$ via Lemma~\ref{lem:ScalingDesignMatrix} results in a matrix $\mB$ with $\|\mB_{i,\bullet}\|_2^2 \leq (1+\varepsilon)\frac{q}{k}$
  and $(1-\varepsilon) \leq \|\mB_{\bullet,j}\|_2^2 \leq (1+\varepsilon)$ for all $i,j$ where we can choose $\varepsilon > 0$ as small as desired. Now consider the positive semidefinite matrix
  $\mM := \mB^\ast\mB$ and note that $\rank(\mA) = \rank(\mB) = \rank(\mM)$.  
  We make use of Lemma~\ref{lem:OffdiagonalEntriesOfDesignMatrix} with parameter $\alpha := (1+\varepsilon)\frac{q}{k}$ to bound the off-diagonal entries in 
  \begin{eqnarray*}
    \|\mM\|_F^2 &=& \sum_{j=1}^n |\mM_{jj}|^2 + \sum_{j_1 \neq j_2} |\mM_{j_1,j_2}|^2 \\ &\stackrel{\textrm{Lem~\ref{lem:OffdiagonalEntriesOfDesignMatrix}}}{\leq}& \sum_{j=1}^n \|\mB_{\bullet,j}\|_2^4 + t(1+\varepsilon)\frac{q}{k}\Big(1-\frac{1}{q}\Big) \|\mB\|_F^2 \\
    &\leq& (1+\varepsilon)(1 + \frac{t(q-1)}{k}) \|\mB\|_F^2, 
  \end{eqnarray*}
 where the last inequality follows by $\sum_{j=1}^n \|\mB_{\bullet,j}\|_2^2
= \|\mB\|_F^2$ and $\|\mB_{\bullet,j}\|_2^2 \leq 1+\varepsilon$ for all $j$. Recall that $\Tr[\mM] = \|\mB\|_F^2 = \sum_{j=1}^n \|\mB_{\bullet,j}\|_2^2 \geq (1-\varepsilon)n $. Then
  \begin{eqnarray*}
   \rank(\mM) &\geq& \left\lceil \frac{\Tr[\mM]^2}{\|\mM\|_F^2} \right\rceil \geq \left\lceil \frac{\|\mB\|_F^2}{(1+\varepsilon)(1+\frac{t(q-1)}{k})} \right\rceil \\ &=& \left\lceil \frac{(1-\varepsilon)}{(1+\varepsilon)}\frac{n}{(1+\frac{t(q-1)}{k})} \right \rceil \geq \frac{n}{(1+\frac{t(q-1)}{k})} \\ &>& n - \frac{nt(q-1)}{k}. 
\end{eqnarray*}
where the second to last step follows by choosing $\varepsilon > 0$ small
enough, and the last step follows from $\frac{1}{1+x} > 1-x$ for all $x > 0$.
\end{proof}

\section{A proof of Theorem~\ref{thm:SGConfigLinesLemmaImproved}}\label{sec:bound-maxlines}

Finally we prove that $f(d) \geq 1-\frac{4}{d+1}$. Once we have Theorem~\ref{thm:RankOfDesignMatrix}, the proof of Theorem~\ref{thm:SGConfigLinesLemmaImproved} is analogous to \cite{dvir2014improved}
and we include it only for completeness. Fix a set of points $S \subseteq \mathbb{C}^d$. We call a
line $\ell \in \textrm{lines}(S)$ \emph{ordinary} if it contains exactly two points from $S$, i.e.
$|\ell \cap S|=2$. In contrast we call the line $\ell$ \emph{special} if it contains more than two points
from $S$. In matroid literature, \emph{special} lines are sometimes denoted by \emph{long} lines. 
\begin{figure} \label{fig:SpecialVsOrdLines}
 \begin{center}
  \includegraphics{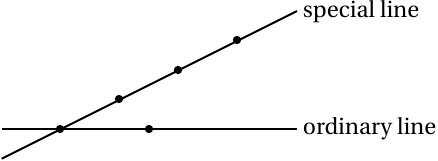} 
\end{center}
  \caption{Special vs. ordinary lines}
\end{figure}
We can paraphrase Lemma~5.1 from \cite{dvir2014improved} as follows:
\begin{lemma} \label{lem:ExistenceOfAffineDepMatrixA}
  Let $S = \{ v_1,\ldots,v_n\} \subseteq \mathbb{C}^d$ be a set of points so that for each $v_i \in S$, at least $k$ % $k := \lceil \delta (|S|-1) \rceil$
  many points in $S \setminus \{ v_i\}$ lie on special lines through $v_i$. Let $\mV \in \mathbb{C}^{n \times d}$ be the matrix with row vectors $v_1,\ldots,v_n$. Then there is a $(3,3k,6)$-design matrix $\mA \in \mathbb{C}^{m \times n}$ with $\mA \begin{pmatrix} \mathbf{1}_n ~ \mV \end{pmatrix} = \bm{0}$.
\end{lemma}
The rows of the matrix $\mA$ are a subset of all the minimal affine dependencies that arise from the special lines. Again, for a proof see \cite{dvir2014improved}. Now we will use this construction to prove Theorem~\ref{thm:SGConfigLinesLemmaImproved}.
\begin{proof}[Proof of Theorem~\ref{thm:SGConfigLinesLemmaImproved}]

We fix a dimension $d \geq 5$, noting that $1-\frac{4}{d+1} > 0$; the goal is to show that $f(d) \geq
1-\frac{4}{d+1}$. Let $S = \{v_1,\ldots,v_n\} \subseteq \CCC^d$ be any set of
points satisfying $\aff(S) = \CCC^d$~\footnote{One can reduce dimension while
maintaining all affine dependencies until this is satisfied using standard
arguments.} and let $\maxlines(S) = r_S \cdot n \leq n-1$ for some $r_S \in
\QQ$. We wish to show that $r_S \geq 1-\frac{4}{d}$. Note that $n \geq d+1$ by
assumption that $\aff(S) = \CCC^d$. Therefore, if $\maxlines(S) = |S|-1 = n-1$,
then $r_S = 1-\frac{1}{n} \geq 1-\frac{1}{d+1}$, and the statement trivially
holds for $S$. Therefore, we may assume that $\maxlines(S) < n-1$. 
   
Take any point $v_i \in S$. Let $\ell_i$, $\ell_i'$ denote the number of lines
and special lines through $v_i$ respectively, and let $k_i$ denote the number
points on special lines through $v_i$. Noting the identity $n-1 - \ell_i +
\ell_i' = k_i$, and the fact that $\ell_i' \geq 1$ by assumption
that $\maxlines(S) < n-1$, we conclude that $k_i \geq n-\ell_i \geq (1-r_S)n$.  

Therefore, letting $k := (1-r_S)n$ we have that Lemma~\ref{lem:ExistenceOfAffineDepMatrixA} applies and that there is a $(3,3k,6)$-design matrix $\mA \in \mathbb{C}^{m \times n}$ so that $\mA \begin{pmatrix} \mathbf{1}_n ~ \mV \end{pmatrix} = \bm{0}$ where $\mV := (v_1,\ldots,v_n)^\T$. Applying the rank lower bound for design matrices from Theorem~\ref{thm:RankOfDesignMatrix} gives
\[
\rank(\mA) > n - \frac{6(3-1)n}{3k} = 1+\frac{4n}{(1-r_S)n} = n - \frac{4}{(1-r_S)}.
\]
Since $\aff(S) = \CCC^d$, note that the column span of $\mV$ cannot contain
$1_n$, since otherwise the points in $S$ would be contained in a hyperplane
(and hence could not have affine span $\CCC^d$). In particular,
$\rank\begin{pmatrix} \mathbf{1}_n ~ \mV \end{pmatrix} = 1 + \rank(\mV) = d+1$. Using
that the columns of $\begin{pmatrix} \mathbf{1}_n ~ \mV \end{pmatrix}$ are in the kernel of $\mA$, by
the rank-nullity theorem we have that $n \geq \rank(\mA) + (d+1)$. In
particular, $d+1 \leq \frac{4}{(1-r_S)}$. Rearranging gives $r_S \geq
1-\frac{4}{d+1}$. This gives the desired bound.
\end{proof}

\section{Applications of the column bound}
\subsection*{Ordinary flats of a point configuration}
Our column bound can be used to bound the number of ordinary flats in a point configuration as in \cite{dvir2014improved}. A $k$-dimensional flat of a point configuration is a rank $k$ minor of the affine matroid of the point configuration. An ordinary flat is then a flat of the form $H \cup \{v\}$ where $H$ is a rank $(k-1)$ flat and $H \cup \{v\}$ is a rank $k$ flat. In the results of \cite{dvir2014improved}, the number of ordinary rank $k$ flats containing a rank $(k-1)$ flat is bounded for any rank $d$ point configuration. In \cite{geelen2024sylvester}, the average number of ordinary rank $k$ flats of a rank $d$ point configuration is bounded. Our analysis can similarly be leveraged to find an $\epsilon > 0$ and $k \geq 1$ such that there exists a rank $(k-1)$ flat contained in $\epsilon \cdot n$ many ordinary $k$ flats. 

In particular, by definition of $f$, this holds for any $\epsilon \leq (2\gamma - 1)$ where $\gamma = \prod_{i=1}^{k-1} f(d-i)$. 

\cortwo*

We prove \Cref{cor: inductiveClaim} by leveraging our bound on the size of a minor free matroid in \Cref{lem:find-minor}. 

\begin{proof}[Proof of \Cref{cor: inductiveClaim}]
   For any independent set $\{e_1, \hdots, e_{k-1}\}$ of $\mathcal{M}$, the flat $flat(e_1, \hdots, e_{k-1}) \cup \{v\}$ is ordinary in $\mathcal{M}$ if and only if $v$ is not parallel with any element in $\mathcal{M'} = \mathcal{M} / \{e_1, \hdots, e_{k-1}\}$. As being parallel is an equivalence relation, there are at most $|\mathcal{M}'| - |si(\mathcal{M}')|$ elements of $si(\mathcal{M})$ which had a parallel element in $\mathcal{M'}$. Thus the number of ordinary $k$ flats containing $flat(e_1, \hdots, e_{k-1})$ is given by at least $|si(\mathcal{M}')|-(|\mathcal{M}'|-|si(\mathcal{M}')|)$. \Cref{lem:find-minor} shows that for any $j$, $\cM$ contains a rank $j$ minor with at least $n \prod_{r=j}^{d-1} f(r)$ non-collinear elements. Thus there exists an independent set $\{e_1, \hdots, e_{k-1}\}$ such that $\cM' = \cM / \{e_1, \hdots, e_{k-1}\}$ is a rank $d-k+1$ minor that contains $n \prod_{r=(d-k+1)}^{d-1} f(r)$ non-collinear elements. Then $$|si(\mathcal{M}')| \geq n \prod_{r = (d-k +1)}^{d-1}f(r) = n \prod_{i=1}^{k-1}f(d-i)$$ such that $|si(\mathcal{M}')|-(|\mathcal{M}'|-|si(\mathcal{M}')|) \geq (2\prod_{i=1}^{k-1}f(d-i) - 1)n$. Letting $\epsilon = 2\prod_{i=1}^{k-1}f(d-i) - 1$, this shows that there are at least $\epsilon n$ ordinary $k$-flats passing through a $k-1$-flat, $H =flat(e_1, \hdots, e_{k-1})$. In particular, using the bound $f(d) \geq (1 - \frac{4}{d+1})$ from \Cref{thm:SGConfigLinesLemmaImproved}, we get that for any $\epsilon > 0$, when $ 2\left( 1 - \frac{k-1}{d}\right)\left( 1 - \frac{k-1}{d-1}\right)\left( 1 - \frac{k-1}{d-2}\right)\left( 1 - \frac{k-1}{d-3}\right) - 1 \geq \epsilon$, there exists $\epsilon n$ many $k$-flats passing through a $k-1$-flat. Using the inequalities \begin{align*}\left( 1 - \frac{k-1}{d}\right)\left( 1 - \frac{k-1}{d-1}\right)\left( 1 - \frac{k-1}{d-2}\right)\left( 1 - \frac{k-1}{d-3}\right) - 1 & \geq \left( 1 - \frac{k-1}{d-3}\right)^4 \\ & \geq \left(1 - \frac{4(k-1)}{(d-3)}\right)\end{align*} then for any $k$ where $\left(1 - \frac{4(k-1)}{(d-3)}\right) \geq \frac{1 + \epsilon}{2}$, the conditions are satisfied and we conclude that this holds when $k \leq \frac{(d-3)(1-\epsilon)}{8}+1$. 
    
\end{proof}

The above proof uses $\maxlines$ estimates defined via the function $f$, which is a bound on the number of distinct lines. This provides a simplified proof on the number of ordinary flats of a rank $d$ point configuration. \Cref{cor: inductiveClaim} essentially recovers the bounds shown in \cite{dvir2014improved}, 
stated in the converse and through an alternative induction method. The main qualitative difference here is that our method inducts over all distinct lines rather than only over ordinary lines. This allows the induction to proceed to dimension 2 using low dimensional bounds on the number of distinct lines, whereas in \cite{dvir2014improved} the induction over ordinary lines is halted at a higher dimension.

\subsection*{Proximity bounds in integer programming}
We next show that \Cref{thm:main-kappa} implies \Cref{cor: graverBasisCor} which gives a bound on the proximity of an integer solution of an integer program to the optimal solution of its LP relaxation in terms of the size of the Graver basis elements. Such proximity bounds are powerful tools in theoretical analysis of integer programs.

\corone*

To prove the proximity bound, we will first rewrite $IP$ into a separable convex integer program with distinct columns in the constraint matrix. That is, an integer optimization problem where the objective function is a separable piecewise linear function. We will also bound the collinearities of the constraint matrix to get a matrix of distinct non-collinear columns for which we can apply the column bound of \Cref{thm:main-kappa}. Our main technical lemma bounds the number of distinct columns of a rational matrix $A \in \mathbb{Q}^{d \times n}$ which will then be used to bound the proximity.

\begin{lemma}\label{lem: number of distinct columns}
    For any rational matrix $A \in \mathbb{Q}^{d \times n}$ where columns of $A$ are all distinct, the number of columns $n \leq d^4 \kappa_A g_\infty(A)^2$. Moreover, $\kappa_A \leq g_\infty(A)$ such that $n \leq d^4 g_\infty(A)^3$.  
\end{lemma}

\begin{proof}[Proof of \Cref{lem: number of distinct columns}]
    \Cref{thm:main-kappa} allows us to bound the number of distinct non-collinear columns of a matrix $A$, so in order to derive the bound on distinct columns, we must handle the collinearities in $A$. If $A$ is comprised of distinct columns but has collinearities, we can assume an ordering on the columns of $A$ such that $$A = \begin{pmatrix}
        A_{Q_1} &A_{Q_2} & \hdots & A_{Q_w} \\
        \vdots&\vdots& \ddots & \vdots
    \end{pmatrix}$$ 
    and each block $A_{Q_i}$ is a set of collinear columns of $A$. Let $k$ be the maximum number of columns per block and let the first block $A_{Q_1}$ contain exactly $k$ columns. Then $n \leq k \cdot w$. Let $A'$ be a matrix containing one arbitrary representative column per block. Then $A'$ is a matrix comprised of $w$ distinct non-collinear columns so that by applying \Cref{thm:main-kappa}, we know that $w \leq \pi d^4 \kappa_{A'}$.  
    
     It remains only to give a bound on $k$ the number of collinear columns in any block of $A$ to conclude. We claim that $k \leq O(g_\infty(A_{Q_1})^2)$.

    To see this, we let $\hat{a}$ be any rational vector such that $$A_{Q_1} = \begin{pmatrix}
        z_1 \hat{a} & z_2 \hat{a} & \hdots & z_k\hat{a} \\
        \vdots & \vdots & \ddots & \vdots 
    \end{pmatrix}$$
    for $z_1, \hdots, z_k \in \ZZ$ distinct integers. Note that this must exist as $A_{Q_1}$ is made up of distinct, collinear rational columns. Assume without loss of generality that $z_1 < z_2 < \hdots < z_k$. If any pair of integers $z_i = -z_j$, remove $-z_j$ from the list such that we have a list of at least $k/2$ many integers distinct in absolute value and reindex these to get a list $z_1 < z_2 < \hdots < z_{k/2}$. Notice that for every $i \in [2, k/2]$, the vector $v^{(i)}$ defined by $$ v^{(i)}_j = \begin{cases}\frac{z_1}{gcd(z_1, z_i)} & \text{ if } j =1\\   -\frac{z_i}{gcd(z_1, z_i)} & \text{ if } j = i \\  0 & \text{ if } j \notin \{1, i\}\end{cases}$$ is in the Graver basis of $A_{Q_1}$ since $A_{Q_1}v^{(i)} = 0$ and minimality is given by the fact that the kernel vector has support size $2$ with no divisors. Then we claim that there exists some $i \in [2, k/2]$ such that $\Vert v^{(i)}\Vert_\infty \geq \sqrt{k/2-1}$. Assume this were not the case so that for every $i \in [2, k/2]$, both $\left|\frac{z_i}{gcd(z_1, z_i)}\right| < \sqrt{k/2-1}$ and $\left|\frac{z_1}{gcd(z_1, z_i)}\right| < \sqrt{k/2-1}$. Then, we have $k/2 -1$ many integer tuples of the form $\left(\left|\frac{z_1}{gcd(z_1, z_i)}\right|, \left|\frac{z_i}{gcd(z_1, z_i)}\right| \right)$ with components bounded between $0$ and $\sqrt{k/2 -1}$. By the pigeonhole principle, there must then exist two identical tuples $$\left(\left|\frac{z_1}{gcd(z_1, z_i)}\right|, \left|\frac{z_i}{gcd(z_1, z_i)}\right| \right) = \left(\left|\frac{z_1}{gcd(z_1, z_j)}\right|, \left|\frac{z_j}{gcd(z_1, z_j)}\right| \right).$$
    But then, $\left|\frac{z_1}{gcd(z_1, z_i)}\right| = \left|\frac{z_1}{gcd(z_1, z_j)}\right|$ implies that $\left|gcd(z_1, z_i)\right| = \left|gcd(z_1, z_j)\right|$ and $\left|\frac{z_i}{gcd(z_1, z_i)}\right| = \left|\frac{z_j}{gcd(z_1, z_j)}\right|$ implies that $|z_i| = |z_j|$. This contradicts the distinctness of $z_1, \hdots, z_{k/2}$ in absolute value. Thus there exists some $v^{(i)}$ in $\mathcal{G}({A_{Q_1}})$ such that $\Vert v^{(i)}\Vert_\infty \geq \sqrt{k/2-1}$. Moreover, as $A_{Q_1}$ is a submatrix of columns of $A$, $\mathcal{G}({A_{Q_1}}) \subseteq \mathcal{G}(A)$ so that $k \leq 2 \Vert v^{(i)}\Vert_\infty^2 + 2 \leq O(g_\infty(A_{Q_1})^2) \leq O(g_\infty(A)^2)$. This concludes the first part of the lemma with $n \leq k \cdot w \leq O(d^4 g_\infty^2(A)\kappa_{A'}) \leq O(d^4 g_\infty^2(A)\kappa_{A})$.

    Finally, we show the second part of the lemma by showing that $\kappa_{A'}\leq g_\infty(A')$.
    To prove this, let $x \in \RR^n$ be a minimal linear dependence of $A'$ such that  $\frac{|x_i|}{|x_j|} = \kappa_{A'}$ for some $i, j \in [n]$. Note that since $A'$ is rational, the kernel of $A'$ is generated by a set of rational vectors, and by scaling $x$ by the least common denominator of all components we do not change the value $\frac{|x_i|}{|x_j|}$, such that we may assume $x \in \ZZ^n$. Since $x$ is an integral minimal linear dependence, if $x$ is not in the Graver basis $\mathcal{G}(A')$, then there must exist some $y \in \mathcal{G}(A')$ such that $\supp(y) \subseteq \supp(x)$, $y$ is sign compatible with $x$, and $|y| \leq |x|$. As $x$ has minimal support in the kernel, it must be that $\supp(y) = \supp(x)$ and that there exists a component $i$ such that $|y_i| < |x_i|$. But if $x$ and $y$ have the same support and component-wise signs, and are minimal kernel elements of $A'$, then they must be linearly dependent. Then $x = \alpha y$ for some $\alpha > 1$ so that 
    \begin{align*}
        \kappa_{A'}  = \frac{|x_i|}{|x_j|} & = \frac{\alpha|y_i|}{\alpha|y_j|}  \leq g_\infty(A')
    \end{align*}
    where the last inequality comes from $y \in \mathcal{G}({A'})$ and $y$ integral such that $|y_j| \geq 1$. The lemma follows as $g_\infty(A') \leq g_\infty(A)$.
\end{proof}

 Using this \Cref{lem: number of distinct columns}, we can now conclude \Cref{cor: graverBasisCor} and give a proximity bound.

\begin{proof}[Proof of \Cref{cor: graverBasisCor}]
    To prove the proximity bound, we first show that we may reduce the problem to a separable convex integer program where the constraint matrix is comprised of distinct columns. 

    To show this, we take the integer matrix $A \in \ZZ^{d \times n}$ and partition $[n]$ into index sets $J_1, \hdots, J_r$ such that for each $i \in [r]$, $J_i = \{i_1, \hdots, i_{q_i}\}$ and $a_{i_j} = a_{i_k}$ for every column index $j, k \in i_q$. That is, $J_1, \hdots, J_r$ is a partition of the column indices into sets of distinct column vectors. Next let $a_{i_1}$ be a representative column vector for the index set $J_i$ for each $i \in [r]$. Let $A'$ be defined as the matrix of the representative columns: $A' := \begin{pmatrix}
                a_{1_1} & a_{2_1} & \hdots & a_{r_1} \\
                \vdots & \vdots & \ddots & \vdots 
            \end{pmatrix}$. 
    This matrix $A'$ will be the constraint matrix for our reduced optimization problem. We will assume without loss of generality by reordering that for each $i \in [r]$, $c_{i_1} \leq c_{i_2} \leq \hdots \leq c_{i_{q_i}}$ such that the cost vector of $IP$ is ordered by decreasing weights over the repeated columns. For each $i \in [r]$, define the partial sum $S_i(k)$ by $S_i(k) = \displaystyle\sum_{j = 1}^k u_{i_j}$ for any $k \in [q_i]$. Then for each $i \in [r]$, define the following piecewise linear function: $$f_i(x): = \begin{cases}
                c_{i_1}x & \text{ if } x \in [0, S_i(1)] \\
                c_{i_1}u_{i_1} + c_{i_2}(x- u_{i_1}) & \text{ if } x \in (S_i(1), S_i(2)] \\
                
                c_{i_1}u_{i_1} + c_{i_2}u_{i_2} + c_{i_3}(x- (u_{i_1} + u_{i_2})) & \text{ if } x \in (S_i(2), S_i(3)] \\
                \vdots \\
                \sum_{j=1}^{q_i -1}c_{i_j}u_{i_j} + c_{i_{q_i}}(x- \sum_{j=1}^{q_i -1}c_{i_j}u_{i_j}) & \text{ if } x \in (S_{i}(q_i-1), S_{i}(q_i)]
            \end{cases}$$

            Finally let the optimization problem $IP'$ be given by $$IP' := \min \{\displaystyle\sum_{i =1}^r f_i(y_i) : A'y = b, y \in [0, u'], y \in \ZZ^r\}$$
        where $u'_i = \displaystyle\sum_{j \in J_i}u_j$ for each $i \in [r]$. 

        We claim that this integer optimization program $IP'$ and its fractional relaxation $LP'$ satisfies $\Vert x^{IP}- x^{LP}\Vert_\infty\leq \Vert x^{IP'} - x^{LP'}\Vert_\infty$ and $g_\infty(A') \leq g_\infty(A)$. To show this, first we note that $\mathcal{G}(A') \subseteq \mathcal{G}(A)$. This is as $A'$ is a submatrix of columns in $A$ so that any kernel vector of $A'$ is also a kernel vector of $A$ padded with zeros where columns were removed. This implies that $g_\infty(A) \geq g_\infty(A')$.

        To show that $\Vert x^{IP}- x^{LP}\Vert_\infty\leq \Vert x^{IP'} - x^{LP'}\Vert_\infty$, we will define a function that maps a solution of $LP'$ to some solution for $LP$. For any feasible vector $x$ to the fractional relaxation $LP'$, define a vector $g(x) \in \RR^n$ where for each $i \in [r]$ and each index $i_j \in J_i$ 
        $$g(x)_{i_j} = \begin{cases}
            0 & \text{ if } x_i \leq S_i(j-1) \\
            (x_i - S_i(j-1)) & \text{ if } x_i \in (S_i(j-1), S_i(j)] \\
            u_{i_j} & \text{ if } x_i > S_i(j)
        \end{cases}$$
        where we use the convention $u_{i_0} = 0$ for each $i \in [r]$. 
        % Define also the map that maps a solution of $LP$ to a solution of $LP'$ as follows:
        % $$f(x)_{i} = \begin{cases}
        %     \sum_{j \in J_i} x_{i_j}.
        % \end{cases}$$
        
        Then, let $x^{LP'}$ be an optimal fractional solution of $LP'$ and let $x^{IP'}$ be the closest integer solution of $IP'$ to $x^{LP'}$. By definition of $LP'$ and as the objective $\vec{c}$ is sorted in decreasing order for every index set $J_i$, $g(x^{LP'})$ must be feasible and optimal for $LP$. Thus we can assume that $x^{LP} = g(x^{LP'})$. Moreover, $g(x^{IP'})$ is feasible for $IP$ as all components are integral. Then $$\Vert x^{LP} - x^{IP}\Vert_\infty = \Vert g(x^{LP'}) - x^{IP} \Vert_\infty {\leq \Vert g(x^{LP'}) - g(x^{IP'}) \Vert_\infty}$$ where the last inequality follows from $x^{IP}$ being the closest feasible integer vector and $g(x^{IP'})$ being feasible for $IP$. Lastly, let $i_j$ be an index in $[n]$ at which $\Vert g(x^{LP'}) - g(x^{IP'}) \Vert_\infty = |g(x^{LP'})_{i_j} - g(x^{IP'})_{i_j}|$. For the index $i$, let $j' \in [q_i]$ be the index such that $x_i^{LP'} \in (S_i(j'-1), S_i(j')]$ and likewise let $j'' \in [q_i]$ be the index such that $x_i^{IP'} \in (S_i(j''-1), S_i(j'')]$. Such indices exist uniquely for each $i \in [r]$ by definition of $LP', IP'$ and since $x_i \in [0, S_i(q_i)]$ for any feasible vector $x$ of $LP'$. Assume without loss of generality that $x_i^{LP'} \geq x_i^{IP'}$, as a symmetric argument holds in the other direction. We now show that $| g(x^{LP'})_{i_j} - g(x^{IP'})_{i_j}| \leq \Vert x^{LP'}- x^{IP'}\Vert_\infty$. Towards this end, there are 4 cases for the indices $j', j''$. 
    \begin{paragraph}{Case 1 $j' = j'' = j$:}
            Then $g(x^{LP'})_{i_j} - g(x^{IP'})_{i_j} = x_i^{LP'} - S_i(j-1) - x^{IP'}_i +  S_i(j-1) = x_i^{LP'} - x_i^{IP'}$ where we have used the definition of $g$ in the first inequality. 
        \end{paragraph}
        
        \begin{paragraph}{Case 2 $j' = j, j'' < j$:}
            Then $g(x^{LP'})_{i_j} - g(x^{IP'})_{i_j} = x_i^{LP'} - S_i(j-1) - 0 \leq x_i^{LP'} - x_i^{IP'}$ where we have used the definition of $g$ in the first inequality and the fact that $S_i(j-1) \geq x_i^{IP'}$ when $j'' <j$. 
        \end{paragraph}

        \begin{paragraph}{Case 3 $j' > j, j'' = j$:}
            Then $g(x^{LP'})_{i_j} - g(x^{IP'})_{i_j} = u_{i_j} - (x_i^{IP'} -S_i(j-1) = S_i(j) - x_i^{IP'} \leq x^{LP'}_i - x^{IP'}_i$ where we have used the fact that $S_i(j) \leq x_i^{LP'}$ when $j' >j$ in the last inequality. 
        \end{paragraph}

        \begin{paragraph}{Case 4 $j' > j, j'' < j$:}
            Then $x^{LP'}_i - x^{IP'}_i \geq S_i(j) - S_i(j-1) = u_{i_j} = g(x^{LP'})_{i_j} - g(x^{IP'})_{i_j}$ where we have used the definition of $g$ and the fact that $S_i(j) \leq x_i^{LP'}$ when $j' >j$ and $S_i(j-1) \geq x_i^{IP'}$ when $j'' < j$. 
        \end{paragraph}

        Thus in all 4 cases, $g(x^{LP'})_{i_j} - g(x^{IP'})_{i_j} \leq x^{LP}_i - x^{IP}_i = | x^{LP}_i - x^{IP}_i| \leq \Vert x^{LP'} - x^{IP'}\Vert_\infty$. By a symmetric argument on $g(x^{IP'})_{i_j} - g(x^{LP'})_{i_j}$, we can conclude that $\Vert x^{LP} - x^{IP}\Vert_\infty \leq \Vert x^{LP'} - x^{IP'}\Vert_\infty$.

        Then, we can restrict our attention to separable convex integer programs with constraint matrices $A'$ such that $A'$ is comprised of $r$ many distinct columns. A classic proximity theorem for separable convex integer minimization gives a bound of $\Vert x^{LP'} - x^{IP'}\Vert_\infty \leq  r\cdot g_\infty(A')$ (see e.g. \cite[Theorem 3.1]{hemmecke2014graver}). Then applying the column bound of \Cref{lem: number of distinct columns} for the $r$ many distinct columns of $A'$,  $$\Vert x^{LP'} - x^{IP'}\Vert_\infty \leq r\cdot g_\infty(A') \leq O(d^4 g_\infty^3(A') g_\infty(A')) \leq O(d^4 g_\infty^4(A)).$$
        As we have shown that $\Vert x^{LP} - x^{IP}\Vert_\infty \leq \Vert x^{LP'} - x^{IP'}\Vert_\infty$, we can conclude the corollary. 
\end{proof}

\newpage
\bibliographystyle{siamplain}
\bibliography{biblio}

\end{document}